\documentclass[useAMS,usenatbib]{mnras}
\usepackage{graphicx}
\usepackage{amsmath}
\usepackage{amssymb}
\usepackage[TABTOPCAP]{subfigure}
\usepackage{url}
\usepackage{multirow}
\usepackage{grffile}

\newcommand{\m}{\rm\thinspace m}
\newcommand{\cm}{\rm\thinspace cm}

\newcommand{\cmsq}{\hbox{$\cm^2\,$}}
\newcommand{\msq}{\hbox{$\m^2\,$}}

\newcommand{\s}{\rm\thinspace s}

\newcommand{\Hz}{\rm\thinspace Hz}

\newcommand{\Msun}{\hbox{$\rm\thinspace M_{\odot}$}}

\newcommand{\keV}{\rm\thinspace keV}

\newcommand{\erg}{\rm\thinspace erg}

\newcommand{\ergcmps}{\hbox{$\erg\cm\ps\,$}}

\newcommand{\ps}{\hbox{$\s^{-1}\,$}}

\newcommand{\rg}{\rm\thinspace $r_\mathrm{g}$}

\title[Returning radiation around black holes]{Returning radiation in strong gravity around black holes: Reverberation from the accretion disc}
\author[Wilkins, Garc\'ia, Dauser \& Fabian]{D. R. Wilkins$^{1}$\thanks{E-mail: dan.wilkins@stanford.edu}\thanks{Einstein Fellow}, J. A. Garc\'ia$^{2, 3}$, T. Dauser$^{3}$ and A. C. Fabian$^{4}$\\
$^{1}$Kavli Institute for Particle Astrophysics and Cosmology, Stanford University, 452 Lomita Mall, Stanford, CA 94305, USA \\
$^{2}$Cahill Center for Astronomy and Astrophysics, California Institute of Technology, Pasadena, CA 91125, USA \\
$^{3}$Dr. Karl Remeis-Observatory and Erlangen Centre for Astroparticle Physics, Sternwartstr. 7, 96049 Bamberg, Germany\\
$^{4}$Institute of Astronomy, University of Cambridge, Madingley Road, Cambridge. CB3 0HA, UK \\
}
\begin{document}

\date{Accepted 2020 August 18. Received 2020 August 12; in original form 2020 April 29}

\pagerange{\pageref{firstpage}--\pageref{lastpage}} \pubyear{2020}

\maketitle

\label{firstpage}

\begin{abstract}
We study reflected X-ray emission that returns to the accretion disc in the strong gravitational fields around black holes using General Relativistic ray tracing and radiative transfer calculations. Reflected X-rays that are produced when the inner regions of the disc are illuminated by the corona are subject to strong gravitational light bending, causing up to 47 per cent of the reflected emission to be returned to the disc around a rapidly spinning black hole, depending upon the scale height of the corona. The iron K$\alpha$ line is enhanced relative to the continuum by 25 per cent, and the Compton hump by up to a factor of three. Additional light travel time between primary and secondary reflections increases the reverberation time lag measured in the iron K band by 49 per cent, while the soft X-ray lag is increased by 25 per cent and the Compton hump response time is increased by 60 per cent. Measured samples of X-ray reverberation lags are shown to be consistent with X-rays returning to the accretion disc in strong gravity. Understanding the effects of returning radiation is important in interpreting reverberation observations to probe black holes. Reflected X-rays returning to the disc can be uniquely identified by blueshifted returning iron K line photons that are Compton scattered from the inner disc, producing excess, delayed emission in the 3.5-4.5\keV\ energy range that will be detectable with forthcoming X-ray observatories, representing a unique test of General Relativity in the strong field limit.

\end{abstract}


\section{Introduction}
The reflection of X-rays from the accretion disc provides a unique insight into the extreme environments just outside the event horizons of astrophysical black holes; supermassive black holes accreting in active galactic nuclei (AGN) and stellar mass black holes accreting in Galactic X-ray binaries. Continuum X-rays emitted from a compact corona, composed of energetic particles that are accelerated either from the inner regions of the accretion disc or within the black hole magnetosphere \citep{haardt+91} propagate through the strong gravitational field. A signifiant fraction of the X-rays emitted close to the black hole will be focused towards the event horizon and the inner parts of the accretion disc by strong gravitational light bending.

Those rays incident on the geometrically thin but optically thick accretion disc are reprocessed. A `reflection spectrum' is formed by the combination of Compton scattering, photoelectric absorption, fluorescent line emission and bremsstrahlung emission by the heated plasma \citep{ross_fabian}. The observed reflection from the inner accretion disc is subject to gravitational redshifts as rays propagate out of the deep gravitational potential, and Doppler shifts, owing to the orbital motion of the reflecting material in the disc. This causes `relativistic blurring' of the reflection spectrum, most easily seen in the profile of emission lines, narrow in the frame of the emitter, that become broadened to display a blueshifted peak and an extended redshifted wing arising form the innermost regions \citep{fabian+89}. Broadening is most easily seen in the iron K$\alpha$ fluorescence line at 6.4\keV\ (in the rest frame), owing to the high fluorescent yield of this line \citep{george_fabian} and the fact that it sits relatively isolated in the spectrum such that the broadening can be seen. The broadening of the iron L line along with lines of oxygen, nitrogen and other elements below 1\keV\ blends the lines together, contributing to a soft excess of emission seen above the power law continuum spectrum at these energies.	

In accreting black hole systems where there is little absorption, either due to outflowing winds or more distant obscuration, measurements of the reflected X-rays have begun to reveal a wealth of information about the accretion process. The extremal redshift detected in the redshifted wing of the iron K$\alpha$ fluorescence line reveals the innermost extent of the accretion disc, from which the spin of the black hole can be inferred \citep{brenneman_reynolds}. Measurement of the profile of the broadened wing of the line reveals the pattern of illumination of the accretion disc by the primary X-ray source, or the \textit{emissivity profile} of the disc \citep{1h0707_emis_paper}, in turn probing the extent of the corona over the surface of the accretion disc \citep{understanding_emis_paper} and also the geometry of the accretion flow \citep{taylor_reynolds}.

Extreme variability is observed in the coronal X-ray emission, in particular in a class of AGN, the narrow line Seyfert 1 galaxies, noted for their variability and complex X-ray spectra \citep{gallo_nls1}. The reflected X-rays are seen to respond to changes in the continuum luminosity, though after a time delay, the \textit{reverberation time lag}, corresponding to the additional light travel time between the corona and the reflecting accretion disc \citep{fabian+09,reverb_review}. Observations of this \textit{X-ray reverberation} add a further dimension to the study of accreting black holes. The observed lag times are short \citep{demarco+2012,kara_global}, indicating the compactness of the corona, lying within $2 \sim 9$\rg\ of the disc plane \citep{lag_spectra_paper}. The gravitational radius, $r_\mathrm{g} = GM/c^2$, is the characteristic scale length in the gravitational field and is approximately the radius of the event horizon of a maximally spinning black hole. Measurement of the relative lag times between successive energy bands mirrors the corresponding reflection spectrum, with a longer time delay in energy bands in which there is a greater contribution from the reverberating emission. The redshifted wing of the iron K line, reflected from the inner parts of the disc, closer to the corona, is seen to respond before the rest frame energy of the line, dominated by the outer disc \citep{zoghbi+2012,zoghbi+2013,kara+13,zoghbi+2014}.

The X-rays reflected from the inner disc should be subject to the same gravitational light bending in the strong gravitational field close to the event horizon as the primary rays emitted from the corona. It is therefore expected not only that a significant fraction of the X-rays reflected from the inner disc will be lost within the event horizon, but also that some fraction of these reflected rays will be returned to the disc to be reflected a second time (and indeed further times), which we shall henceforth refer to as \textit{returning radiation}. Returning radiation will impact not just the emissivity profile of the disc, but also the measured reverberation lags as higher order reflections impart additional light travel time delays. Therefore, understanding the effects of returning radiation on X-ray reflection and reverberation observations is an important component of using these as probes of black holes and accretion physics, while direct detection of the effects of returning radiation will validate the prediction of General Relativity that strong light bending should operate in the vicinity of a black hole.

The first studies of returning radiation, conducted by \citet{cunningham-76}, considered its effect on the thermal emission from the accretion disc. It was found that while the effects of returning radiation are negligible when the spin parameter, $a / M < 0.9$, the returning flux can be comparable to or greater than the locally generated flux at small radii ($r < 1.5$\rg) when $a / M > 0.99$, altering not only the thermal emission spectrum but also the dynamics of the disc as returning photons impart a torque. \citet{agol_krolik} incorporated returning radiation into calculations of disc dynamics and found that returning thermal emission becomes more significant when the zero-torque condition is relaxed at the innermost stable orbit (\textit{i.e.} magnetic connections between the plunging region and inner edge of the disc can exert stresses). They find that up to 58 per cent of the thermal emission can return to the the disc. \citet{connors+2020} find that the X-ray reflection spectrum detected in the black hole X-ray binary XTE\,J1550$-$564 during the soft, disc-dominated accretion state is best described by a disc illuminated by a thermal black body spectrum rather than by a power law spectrum from the corona, suggesting that returning radiation is indeed significant and that it is the reflection of returning thermal emission that is detected.

The reflection of returning X-rays was first studied in the context of multiple local reflections on a corrugated accretion disc surface by \citet{ross_fabian_ballantyne}. It was found that when the reflected X-ray emission impacts the disc surface multiple times before being able to escape that the effective ionisation potential is increased, enhancing the observed soft excess, while the iron K$\alpha$ emission line is sharpened and increased by around 20 per cent relative to the continuum.

We extend these studies of returning radiation, considering not just local multiple reflections, but returning radiation in General Relativity between different patches of the disc. Returning rays between distinct patches of the disc undergo gravitational redshift or blueshift and Doppler shifts owing to the relative motion of the disc material at the start and end points of the ray coupled with the direction of the ray. We follow returning rays in the Kerr spacetime using ray tracing simulations and employ radiative transfer calculations to compute the response of the disc to the returning primary reflection. We assess the impact of returning radiation on X-ray reflection spectroscopy and reverberation measurements and assess the detectability of direct signatures of returning radiation in strong gravitational fields.

\section{Tracing the returning radiation}
The passage of photons in the curved spacetime around the black hole was followed using general relativistic ray tracing simulations. Rays were traced by integrating the null geodesic equations in the Kerr spacetime using the \textsc{CUDAKerr} GPU-based ray tracing code \citep{understanding_emis_paper, lag_spectra_paper,propagating_lag_paper,plunging_region_paper}.

\subsection{Primary irradiation}

\begin{figure}
\centering
\includegraphics[width=85mm]{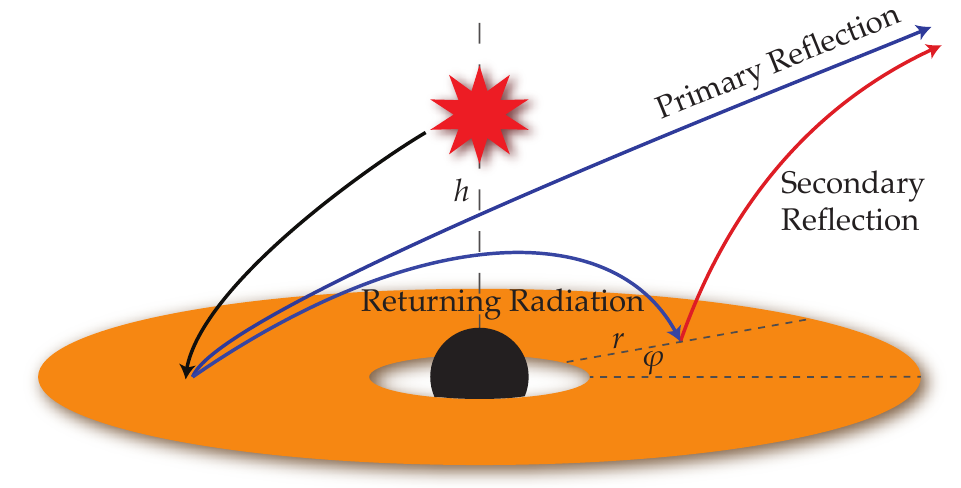}
\caption[]{Schematic of the ray tracing model around the black hole. The X-ray source is assumed to be an isotropically emitting point source located at height $h$ on the rotation axis. Rays emitted from the source are traced, following the null geodesics in the Kerr spacetime, until they reach locations $(r, \varphi)$ the accretion disc in the equatorial plane, where they are reflected. The primary reflection may be observed directly, or reflected rays may return to the disc (the 
\textit{returning radiation}) to produce the secondary reflection component.}
\label{schematic.fig}
\end{figure}

For simplicity, the primary source of the X-ray continuum was taken to be an isotropic point source located at some co-ordinate height $h$ above the singularity on the spin axis of the black hole, illustrated in Fig.~\ref{schematic.fig}. Rays were started at equal intervals in solid angle (starting at equal steps in $\cos\alpha$ and $\beta$, the polar angles) as measured in the source frame. The initial 4-momentum vectors describing each ray were transformed to the global Boyer-Lindquist co-ordinate system (given the 4-velocity of the source) and the constants of motion $h$ and $Q$, corresponding to angular momenta in the $\varphi$ and $\theta$ directions and identifying each ray, were computed. The rays were traced by numerically integrating the first-order geodesic equations, evolving each of the co-ordinates $(t, r, \theta, \varphi)$ as the affine parameter was advanced.

Rays from the primary source were traced until they reached the equatorial plane, in which the geometrically thin, optically thick accretion disc was taken to lie. Upon hitting the disc plane, the time co-ordinate, $t$, of each ray was recorded and the redshift, $g$, of each ray between the source and an observer in a (relativistic) circular orbit at the receiving radius in the disc was calculated following the standard procedure, calculating the ratio of the scalar products of the photon 4-momentum with, respectively, the receiver and emitter's 4-velocity. By assuming the disc is flat, we consider only radiation that returns to the disc by light bending around the black hole. If the disc has a finite aspect ratio, there will be an additional geometric contribution to returning radiation from one side of the disc to the other that we do not consider here.

The equatorial plane was divided into 500 logarithmically-spaced radial bins between the innermost stable circular orbit, representing the inner extent of the accretion disc, at $r=1.235$\rg\ for a maximally rotating black hole with spin parameter $a=0.998\,GMc^{-2}$ (where the spin parameter in the Kerr metric is related to the angular momentum of the black hole by $J = Mac$), and $r = 500$\rg.

The photon arrival rate into the bin at the end of each ray is given simply by the redshift between the source and the corotating observer at the intercept-point on the disc, representing the relative rates at which proper time elapses (since the photon count is conserved along the ray). The flux along each ray is obtained by multiplying by a second factor of the redshift to account for the energy shift of each photon for a monochromatic source, or by $g^\Gamma$ for a source emitting a power law spectrum with the count rate following $E^{-\Gamma}$ \citep[see][]{dauser+12}. The aberration in solid angle between the emitter and receiver is accounted for by counting the relative numbers of rays landing in different bins and does not require additional factors of $g$ in this instance. The total photon count rate and flux incident upon each bin was computed by summing, respectively, the redshift $g$ and $g^2$ for all of the rays landing in the bin, dividing by the total number of rays emitted from the primary source. We average the redshift and arrival time over all of the rays arriving in each bin.

The primary reflection spectrum generated when a power law continuum spectrum irradiates the disc was calculated for each radial bin using the \textsc{xillver} model \citep{garcia+2010,garcia+2011,garcia+2013}, describing the reprocessed X-ray spectrum as emitted in the rest frame of the irradiated material. The ionistation parameter, $\xi = 4\pi F / n$, where $F$ is the ionising flux integrated between 0.1 and 1000\keV, and the density, $n$, were assumed to be constant over the entire disc. For simplicity, these `reflected' photons are assumed to be emitted isotropically in the disc frame. The combination of frame dragging and light bending in the Kerr spacetime brings much of the radiation to the inner disc at glancing angles, leading to limb-brightening that is offset against limb darkening in the disc atmosphere \citep{svoboda+10,garcia+2014}. This, however, is only simply calculated when each patch of the disc is illuminated by just one ray path from a point source. Here, we consider illumination of the disc by rays coming from a range of angles, both from the corona and returning from the disc.

X-rays reflected by the accretion disc are subject to Doppler shifts, due to the orbital motion of the disc material, and gravitational redshifts in the strong field close to the black hole, the combination of which is a function of the position $(r, \varphi)$ on the disc. The energy shift seen from each location on the disc was computed by tracing a regular grid of rays backwards from an image plane a large distance ($r = 10,000$\rg) from the singularity centered on the line-of-sight at angle $\vartheta$ to polar axis, representing the inclination at which the accretion disc is observed. The parallel rays passing through perpendicular to the image plane represent the patch of sky observed by a telescope some large distance away \citep[see][]{propagating_lag_paper,reynolds+99}. The disc is divided into radial and azimuthal bins and the ray travel time and redshift between each bin and the (stationary) image plane was recorded in a lookup table.

\begin{figure}
\centering
\includegraphics[width=85mm]{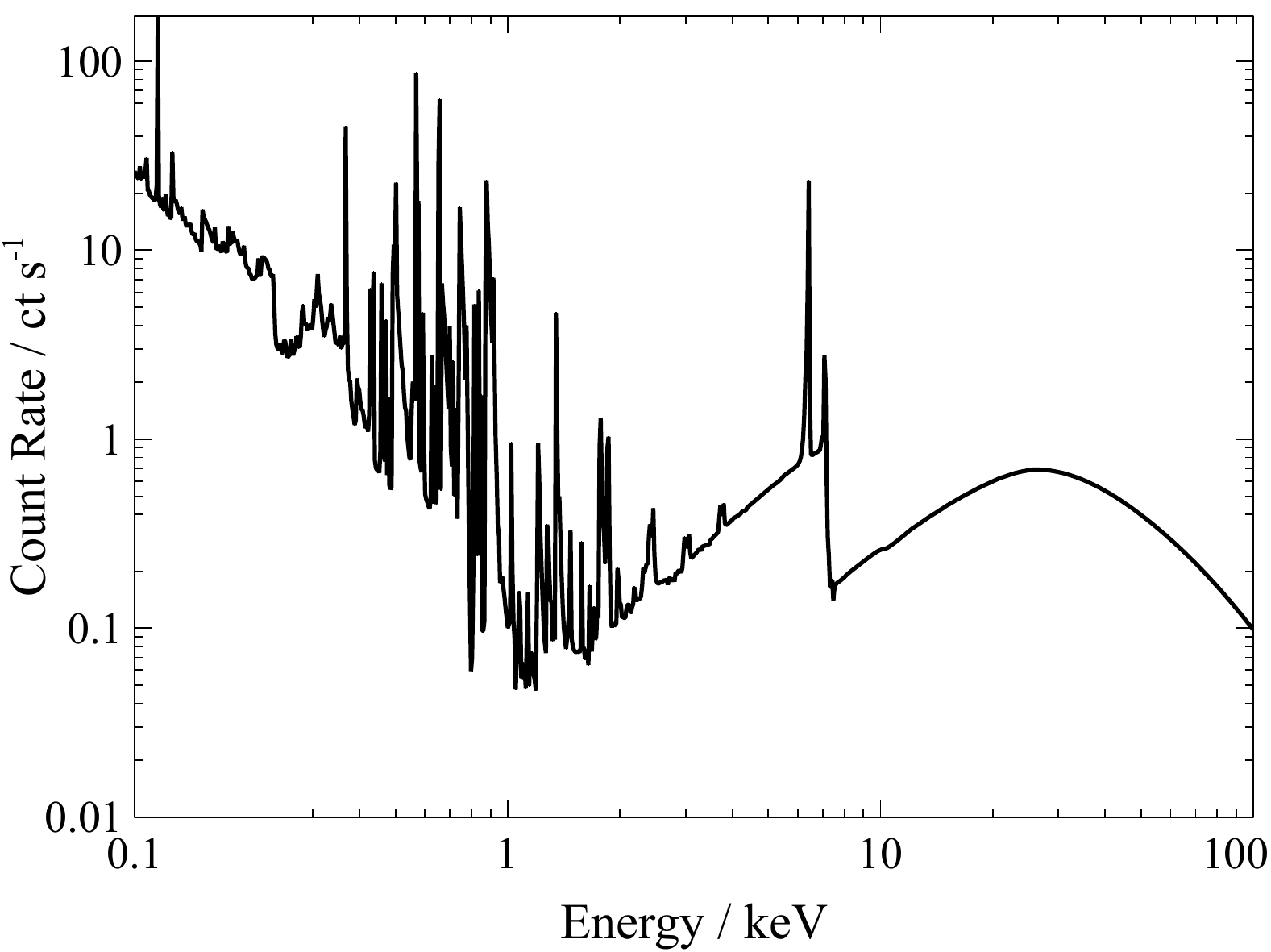}
\caption[]{The \textsc{xillver} model of the primary X-ray reflection spectrum from an accretion disc with iron abundance $A_\mathrm{Fe} = 8$ and ionisation parameter $\xi = 50$\ergcmps\ illuminated by a continuum with photon index $\Gamma = 3$. Returning radiation possesses this spectrum, shifted in energy. The \textsc{xillverRR} model predicts the secondary reflection spectrum emergent from plasma in the accretion disc illuminated by this spectrum.}
\label{xillver.fig}
\end{figure}

When treated separately, the observed primary reflection is computed for each of the radial bins on the disc by iterating through each of the azimuthal bins in the disc-to-observer table. For each radial and azimuthal bin, the rest-frame reflection spectrum described by \textsc{xillver} (Fig.~\ref{xillver.fig}) is shifted in energy according to the bin's redshift and normalised in photon count according to the primary photon count recorded in each bin, and then multiplied by $g^{3}$. Three factors of the redshift, $g$, between the disc and observer account for photon arrival rate recorded by observers at each end of the ray as well as the aberration in solid angle between the source and observer, according to Louiville's Theorem.

\subsection{The returning radiation}
In order to compute the reflected radiation that returns to the accretion disc, a secondary source was placed in the equatorial plane in each radial bin. Due to the axisymmetry of the Kerr metric and illumination of the disc by the primary source, emission from each radial bin can be represented by just a single point source at one azimuthal location. The disc sources were taken to be elements of an isotropically emitting plane, with $dI/d\Omega = \cos\alpha\,d\Omega = \cos\alpha\,d(\cos \alpha) d\beta$, where $\alpha$ is the angle from the normal to the disc and $\beta$ the azimuthal direction of emission (in contrast to the spherically emitting primary source for which $dI / d\Omega = d(\cos \alpha) d\beta$). Disc elements were assumed to be in a stable circular orbit with angular velocity $\Omega = \frac{d\varphi}{dt} = (a^2 + r^\frac{3}{2})^{-1}$.  

Rays were traced from the point source at each radius, integrating the geodesic equations, until they returned to the equatorial plane. Upon returning to the disc, rays were once again counted into radial bins, recording the total returning count rate and flux in each bin as well as the redshift, $g$, and co-ordinate travel time, $t$, of the rays. While the source at each radius emitted the same number of rays to maintain resolution, the count rate and weighting of rays from each source were scaled according to the primary flux that was incident in the corresponding bin to accurately represent the relative contributions of the primary and returning radiation.

The secondary reflection produced by each of the returning rays was modeled using \textsc{xillverRR}, a modification of the \textsc{xillver} model that computes the emergent spectrum when the disc is irradiated using a \textsc{xillver} primary reflection spectrum rather than a power law continuum. The primary reflection spectrum received at each point on the disc, from each other point of the disc, is subject to a different shift in energy, depending upon the relative locations (and hence velocities) of the emitting and receiving positions and the direction of the ray relative to the emitter and receiver. As such, the \textsc{xillverRR} spectrum is computed as a function of the energy shift of the incoming spectrum. Each ray path has only a single energy shift so it is only necessary to shift the energy of the incident spectrum. No blurring kernel needs to be applied at this stage. For each radial bin, the secondary reflection spectrum is computed based upon the average energy shift of all returning rays.

The specific reprocessed spectrum depends upon the total incident spectrum at a given point in time, thus, strictly, the primary and secondary reflection cannot be treated separately. Hence, when considering the time-averaged spectrum that is observed when the primary illuminating flux is constant, the reprocessed spectrum from each position on the disc is computed from the sum of the primary continuum and the primary reflection, meaning that \textsc{xillverRR} has two parameters, the energy shift and the flux ratio between the returning primary reflection and the primary continuum. We here only consider the first order effect of the primary reflection returning to the disc and do not consider the higher orders of returning radiation that will further impact the illuminating and the observed spectrum, but with reducing magnitude.

The observed secondary reflection, or the observed combined reflection spectrum, is calculated in the same manner as the observed primary reflection, exploiting the fact that for illumination by an axisymmetric source, axisymmetry is not broken until the disc is observed from a specific line of sight. Each radial bin is again divided up into azimuthal bins and the lookup table is employed to obtain the energy shift and travel time between each bin and the observer.

\section{Properties of the returning radiation}
General relativistic ray tracing in the Kerr spacetime around the black hole reveals the destinations of reprocessed X-rays that are emitted from the disc; the fraction that return to the disc, to where on the disc they return and the energy shifts and time delays of returning rays when they arrive back at the disc.

\begin{figure}
\centering
\includegraphics[width=85mm]{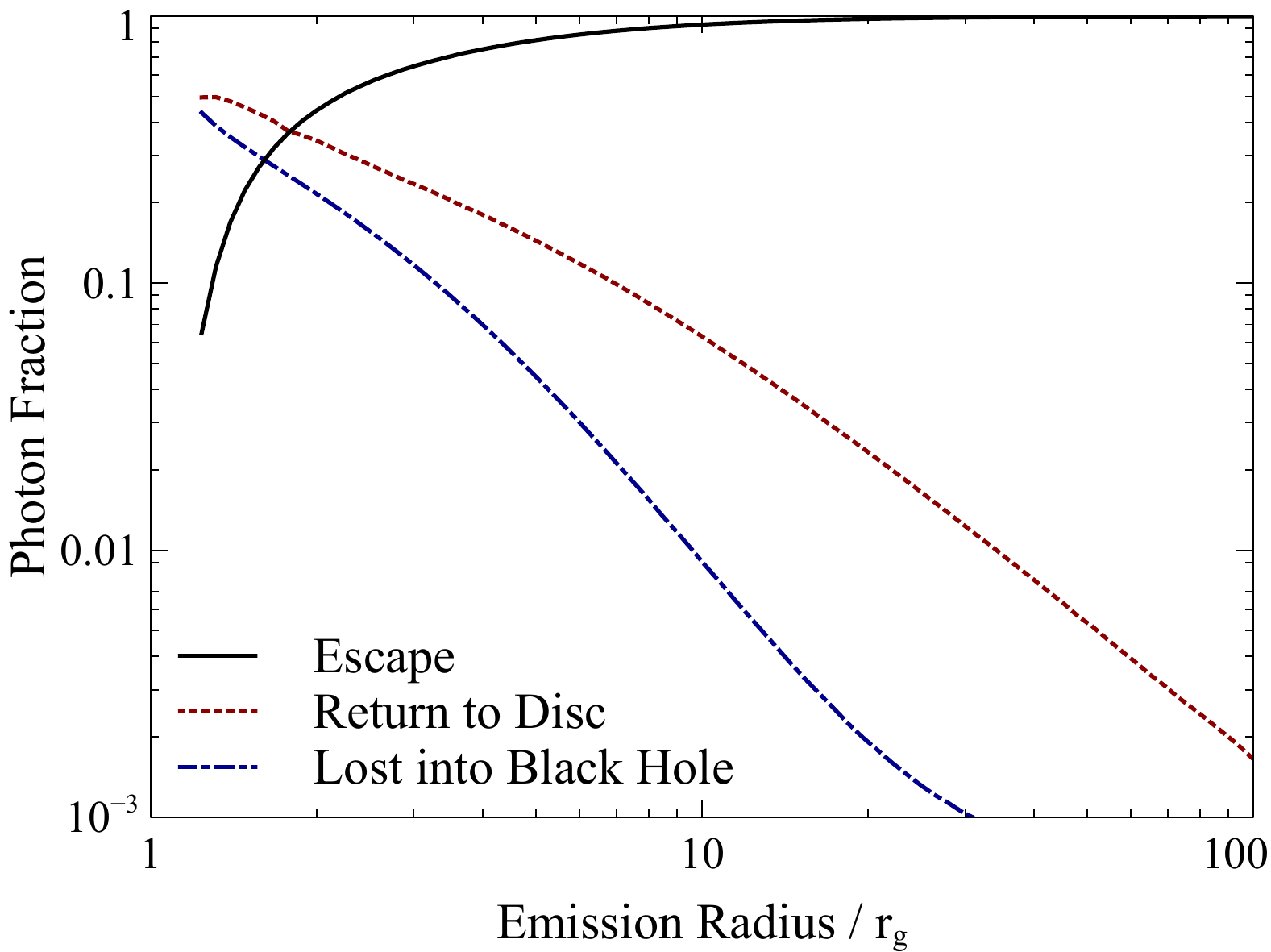}
\caption[]{The fraction of photons emitted isotropically at varying radii on the accretion disc (in the equatorial plane) that are able to escape to be observed as reflected emission, that return to the disc and that are lost inside the event horizon.}
\label{photonfrac.fig}
\end{figure}

Fig.~\ref{photonfrac.fig} shows the fraction of rays that are emitted from each radius on the disc around a maximally spinning ($a=0.998$) black hole that escape (and can be observed as part of the reflection spectrum or reverberation response), defined as rays being able to reach $r=1000$\rg\ without returning to the equatorial plane; the fraction that return to the disc; and the fraction lost within the event horizon of the black hole. The disc is illuminated by a point source at $h=5$\rg\ above the singularity. A greater fraction of rays emitted from the inner parts of the disc pass closer to the black hole, thus are strongly bent in the gravitational field, resulting in a larger fraction of rays from the inner disc returning to the disc.

Up to 50 per cent of the rays emitted from the innermost stable orbit return to the disc, while only 6 per cent are able to escape to be observed. Due to the large solid angle subtended by the event horizon to an observer at the ISCO, up to 43 per cent of the rays emitted will be lost into the black hole. The returning fraction drops substantially to just 12 per cent at 6\rg, the innermost stable orbit for a non-spinning black hole, consistent with the findings of \citet{agol_krolik}. The majority of rays from beyond 10\rg\ are able to escape to be observed in the primary reflection component without being returned to the disc.  Returning radiation is, hence,  much less significant around more slowly spinning black hole, as shown in Table~\ref{spin.tab}, showing the total fraction of rays (emitted from all radii on the disc) that return and escape from the whole disc as a function of black hole spin. We will therefore only consider the maximal spin, $a=0.998$ for the remainder of this work.

\begin{table}
\centering
\caption{\label{spin.tab}The total fraction of rays emitted from the disc that are able to escape (to be observed as part of the X-ray reflection) and that return to the disc, as a function of the dimensionless spin parameter of the black hole. The disc is assumed to be illuminated by an isotropic point source at $h = 5$\rg\ above the singularity.}
 \begin{tabular}{ccc}
 \hline
  Spin, $a / M$ & Escape Fraction, $E_p$ & Return Fraction, $R_p$ \\
  \hline
  0.998 & 0.33 & 0.39 \\
  0.95 & 0.60 & 0.21 \\
  0.9 & 0.66 & 0.18 \\
  0.8 & 0.73 & 0.14 \\
  0.5 & 0.82 & 0.09 \\
  0 & 0.94 & 0.06 \\
  \hline
 \end{tabular}

\end{table}

\begin{figure}
\centering
\subfigure[Primary and returning photon flux] {
\includegraphics[width=85mm]{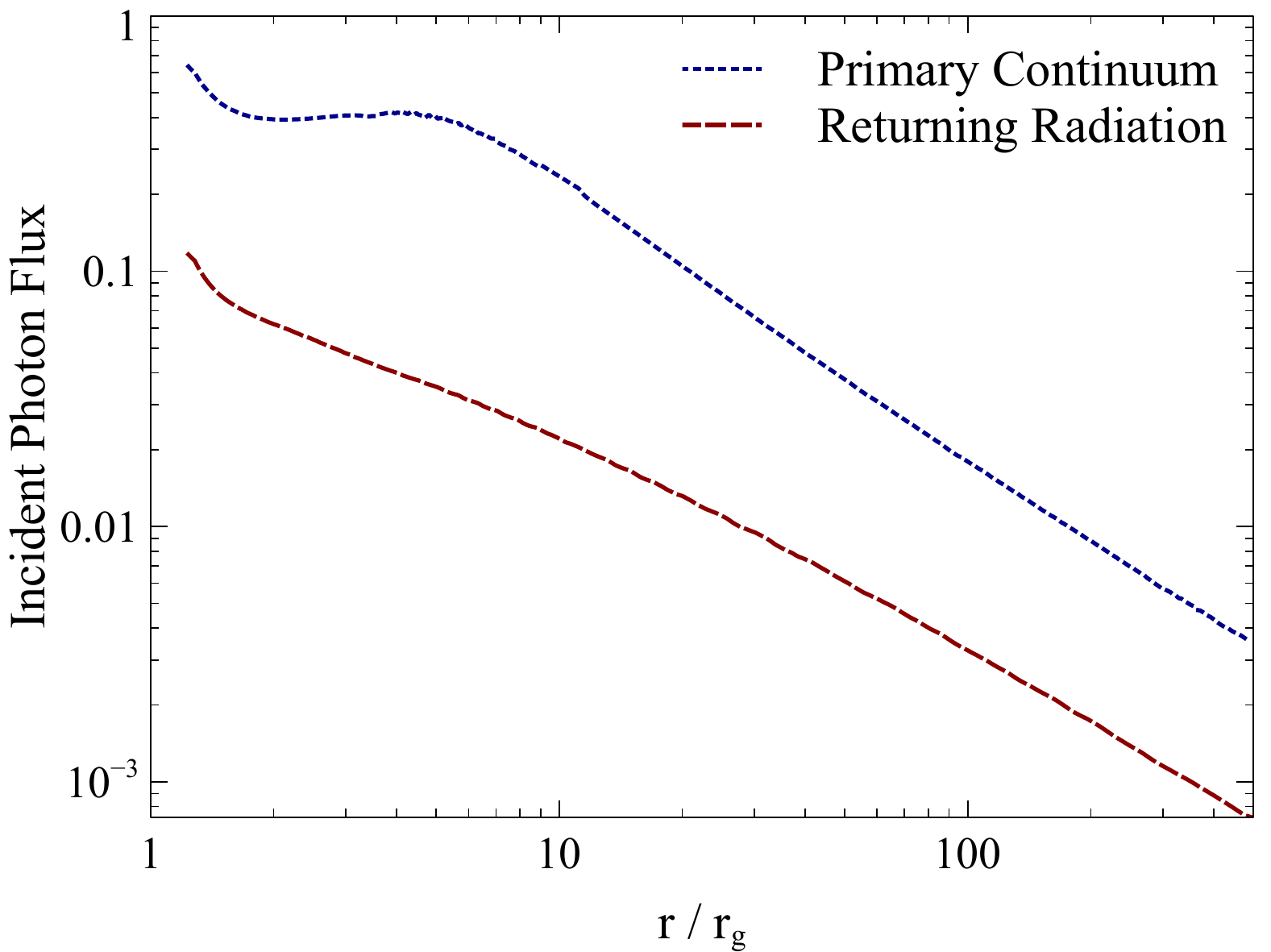}
\label{return_properties.fig:flux}
}
\subfigure[Return to primary photon flux ratio] {
\includegraphics[width=85mm]{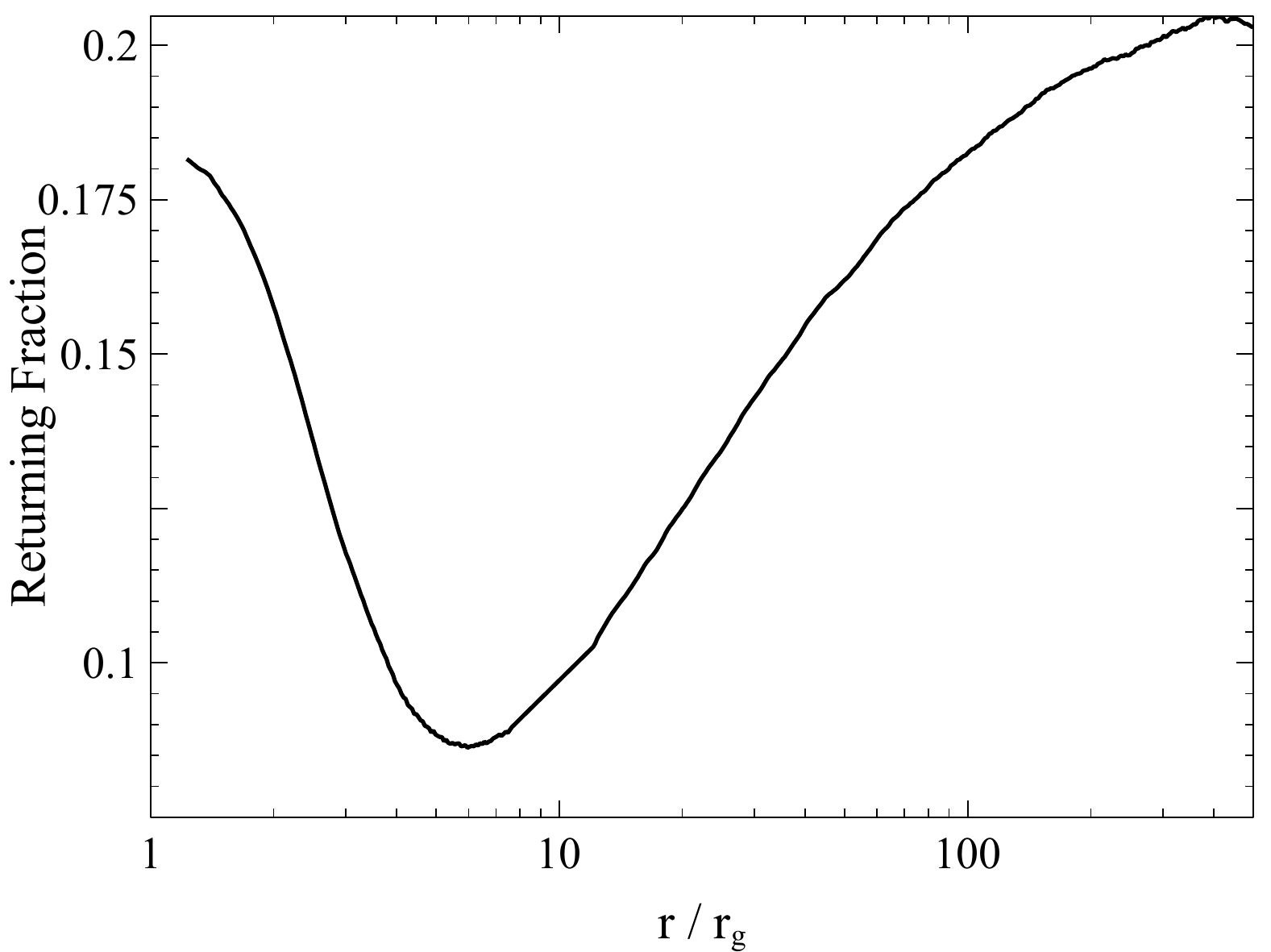}
\label{return_properties.fig:ratio}
}
\subfigure[Energy shift] {
\includegraphics[width=85mm]{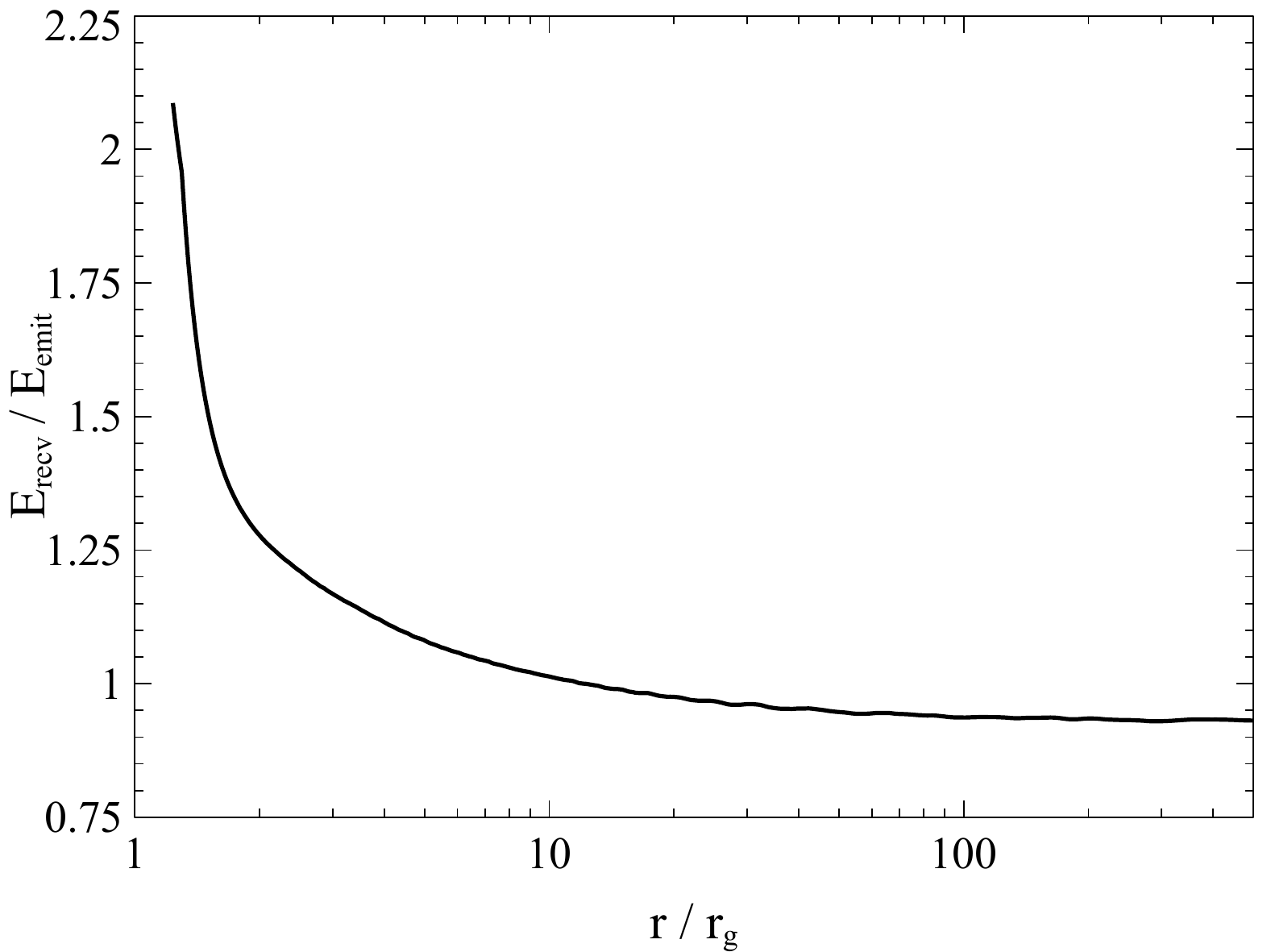}
\label{return_properties.fig:enshift}
}
\caption[]{The properties of the returning radiation received at different radii upon the disc. \subref{return_properties.fig:flux} The returning photon flux, \textit{i.e.} the count rate received per unit area (from the primary reflection) compared to the primary incident photon flux, as measured in the rest frame of the disc. \subref{return_properties.fig:ratio} The ratio of the returning to primary flux. \subref{return_properties.fig:enshift} The average energy shift of returning photons between when they are emitted at the primary reflection site and when they are received upon returning to the disc.}
\label{return_properties.fig}
\end{figure}

Fig.~\ref{return_properties.fig} shows the properties of the returning rays received as a function of radius on the disc. Fig.~\ref{return_properties.fig:flux} shows the illumination pattern as a function of radius of the returning rays compared to the primary illumination by a point source at $h=5$\rg\ above the singularity. Shown is the photon count rate received per unit proper area, measured by an observer co-rotating with the disc and including the effect of length-contraction of the orbiting frame \citep{understanding_emis_paper}. The primary illumination pattern shows the typical broken power law, flattening from $r^{-3}$ over the outer parts of the disc to a flattened profile at radii $r \ll h$. There is a slight steepening on the inner disc due to gravitational light bending focusing the primary emission and the time dilation to observers on the inner disc enhancing the photon arrival rate \citep{understanding_emis_paper}. Shown is the photon arrival rate per unit area, measured in a rest frame co-rotating with the disc, rather than the \textit{emissivity} profile, which is defined as the \textit{energy flux} and is thus enhanced to a steeper power law at small radius by an additional factor of the energy shift between source and disc \citep{dabrowski+97,dabrowski_lasenby,suebsuwong+06,understanding_emis_paper}.

Fig.~\ref{return_properties.fig:ratio} shows the ratio of the returning radiation to the primary illumination as a function of radius. The returning radiation fraction is high on the inner disc, reaching 18 per cent of the primary flux from the corona, dropping to a minimum of 9 per cent at $7$\rg\ and increasing to 20 per cent over the outer disc. The minimum in the flux ratio is due the flattening of the primary emissivity profile where $r \ll h$; the primary irradiation is enhanced with respect to the returning radiation which falls off as $r^{-3}$ over these radii. The location of this minimum is determined by the location of the primary X-ray source. The presence of a minimum in the returning fraction is suggestive of two populations of rays that dominate the return radiation over different parts of the disc. Rays that are emitted from the outer parts of the disc will only return to the disc if they are emitted inwards, towards the black hole, and will return to the inner parts of the disc. Rays emitted from the inner parts of the disc are divided into two populations, depending on their emission direction. Those emitted inwards and counter to the rotation of the disc will return to the disc at small radius, adding to those infalling from further out. On the other hand, rays emitted in the direction of motion from parts of the inner disc can be bent around the black hole to return to the disc at larger radius. It is only this population of rays that will return to the outer parts of the disc.

The returning fraction increases and tends to a constant value at large radius because the rays returning here largely originate from the inner disc. While the returning flux falls off at large radius, so too does the primary flux. At large radii, the majority of the returning radiation, which comes from the inner disc, appears to originate from a point source that adds to the photon count from the primary source. Since the ratio between the returning flux and primary continuum flux at most radii on the disc is less than around 20 per cent, returning radiation does not significantly affect the radial emissivity profile of the accretion disc.

Fig.~\ref{return_properties.fig:enshift} shows the energy shift of the returning radiation. Plotted is the average energy shift of all rays landing in each bin, between their emission measured in the orbiting frame of their source on the disc and their reception, as measured in the receiving frame. The returning energy shifts are weighted by the flux the disc observes returning along each ray. Any individual returning ray could have a range in energy shift due to the varying gravitational redshifts between the points of emission and reception, and the Doppler shift due to the direction of the specific ray relative to the disc. A significant number of the rays returning to the innermost radii on the disc are blueshifted. These rays, in general, originate from larger radius. Rays that are emitted from the disc counter to the direction of motion around a spinning black hole are more likely to be directed towards the event horizon and inner regions of the disc \citep{thorne_evolution}, an effect that is most pronounced for emission radii within $\sim 10$\rg. These rays will be bent around the black hole to encounter orbiting material head-on when they return to the disc, meaning they are Doppler shifted to higher energies, coupled with the blueshift experienced as they travel inwards through the gravitational potential from larger to smaller radii.

On the other hand, rays returning to the outer disc must climb out of the gravitational potential from smaller radii. They are not strongly redshifted, however, because the rays from the inner disc that are able to travel outwards are predominantly those emitted in the forward direction of motion of the emitting material. They are Doppler shifted to higher energy by the orbital motion of the disc before they propagate out of the potential, thus only moderate redshifts are seen in the radiation returning to the outer disc.

\subsection{The reflection fraction}
Enumerating rays emitted from all radii reveals that only 33 per cent of rays that are emitted from the disc around a maximally spinning black hole (\textit{i.e.} 33 per cent of the reprocessed or reflected primary rays) are able to escape to infinity to be observed. This is the primary irradiation escape fraction, $E_p$. 39 per cent of the reprocessed rays return to the disc to be reflected a second time, the primary return fraction, $R_p$.

The total escaping and returning fractions, integrated over the whole disc, are functions of the illumination pattern or emissivity profile of the accretion disc. Greater flux emitted at small radius where the local return fraction is greater will lead to an overall increase in the returning flux. Secondary illumination of the disc follows approximately a constant $r^{-3}$ power law profile (with a slight steepening over the innermost radii), in contrast to the broken power law emissivity profile that is typical of the primary illumination. For this single power law illumination pattern, the escape fraction $E_r = 0.29$ while the return fraction $R_r = 0.41$. For comparison, \citet{agol_krolik} find that up to 58 per cent of the thermal emission returns to the disc, although this is calculated based upon the emissivity profile of the thermal emission from the disc, including a significant flux emitted from the plunging region, inside the ISCO, deep within the gravitational potential. Here we consider the emissivity profile of just the stably orbiting disc due to external irradiation by the corona.

The reflection fraction, here defined as the ratio of the total \textit{observed} reflected flux to the continuum flux, can be used as a measure of the compactness of the corona around the black hole. Emission from a more compact corona, closer to the black hole, is subject to strong light bending, enhancing the flux that irradiates the disc and, hence, enhancing the observed reflected flux at the expense of the directly-observed continuum flux \citep{miniutti+04,mrk335_corona_paper,xmm2015proc}. For a given coronal geometry, the reflection fraction can be estimated from ray tracing calculations by measuring the fraction of rays that reach the disc \textit{vs.} that able to escape to be observed as part of the continuum. The fraction of reflected rays, however, should be adjusted by the fraction of the reflected emission that is able to escape \textit{vs.} that returning to the disc. It should be noted that this definition of the reflection fraction; the ratio of the total number of rays that reaches the disc to that which escapes, differs from the definition of the reflection fraction in \textit{e.g.} \citet{dauser+15}, where the reflection fraction is the ratio of reflected to continuum flux received by an observer at a given inclination.

A fraction $E_p$ of the primary reflection directly escapes to be observed as the reflection component. A fraction $R_p$ return to the disc. In addition to the $E_p$ observed primary reflection, a fraction $E_r$ of the $R_p$ rays will escape in the secondary reflection component. A fraction $R_r$ of the secondary return to the disc again and will form higher order returning radiation components.

Ray tracing calculations show that the while the illumination profile changes from a broken power law to approximately a single $r^{-3}$ power law between the primary and secondary reflection, illumination of the disc by higher order returning components where the disc source flux follows $r^{-3}$ is not substantially altered from this power law, thus the same fractions $E_r$ and $R_r$ may be used for all higher order reflection components.

The total observed reflected flux is obtained by multiplying the primary flux incident upon the disc by the total escape fraction
\begin{align}
\label{gp.equ}
\begin{split}
 E &= E_p + R_p\left(E_r + R_rE_r + R_r^2E_r + ...\right) \\
 &= E_p + R_p E_r \sum_{N=0}^\infty R_r^N
 \end{split}
\end{align}

Summing this geometric progression to infinity yields
\begin{align}
\label{escapefrac.equ}
 E = E_p + R_p \frac{E_r}{1 - R_r}
\end{align}

Hence, for a point source located at \textit{e.g.} $h=5$\rg\ above the singularity, the intrinsic reflection fraction of 2.1 \citep{xmm2015proc} is reduced to 0.52 times that value due to reflected radiation returning to the disc in the strong gravitational field around the black hole.

\begin{figure}
\centering
\includegraphics[width=85mm]{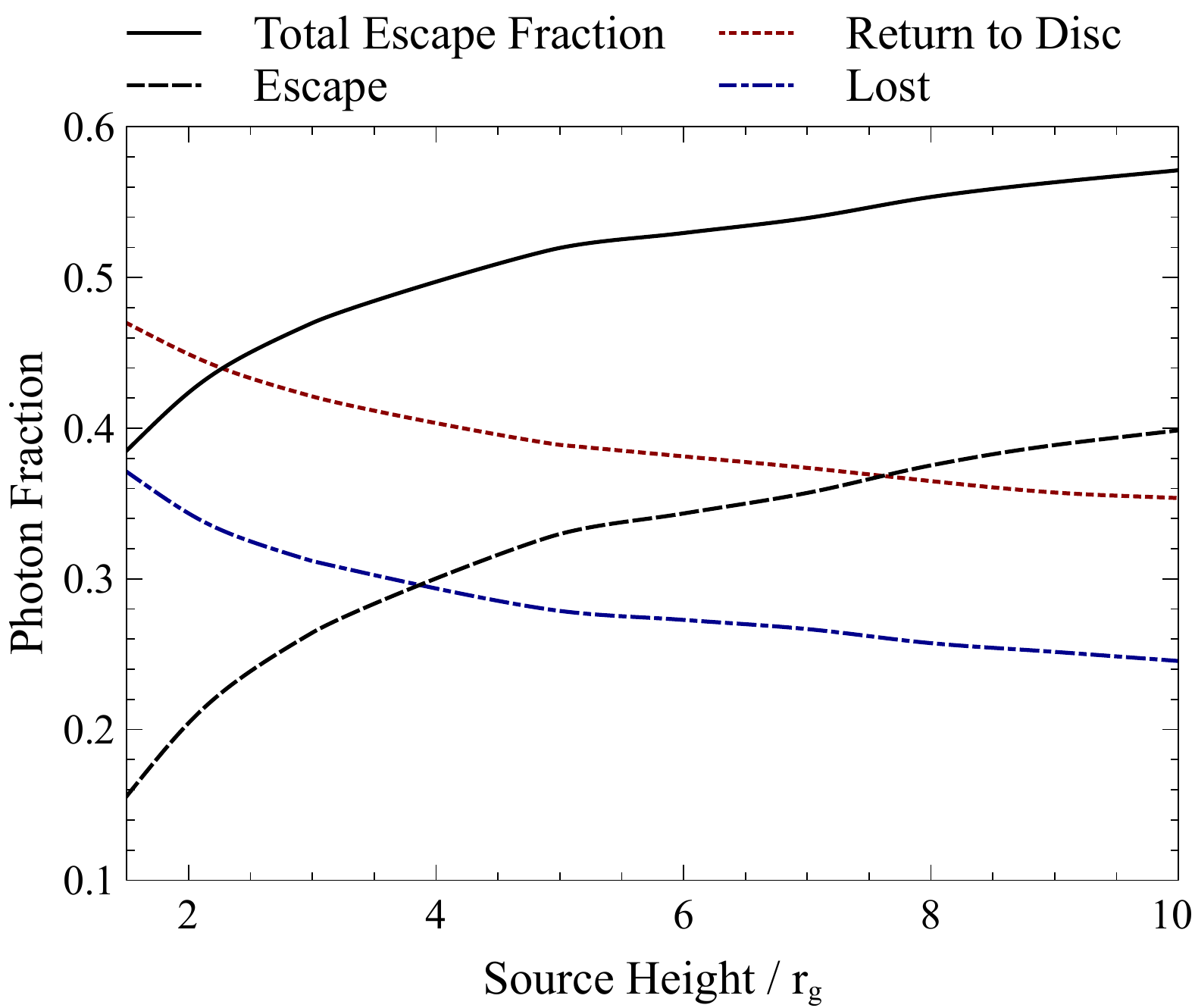}
\caption[]{The fraction of photons emitted from all radii on the disc that are able to escape (to be observed in the reflection component), that return to the disc and are lost into the event horizon as a function of the height of the primary point source above the singularity. The solid line shows the total escape fraction after infinite returns.}
\label{return_fraction_h.fig}
\end{figure}

The returning fraction varies with the geometry and location of the primary X-ray source. For a more compact source, closer to the black hole, a greater fraction of the emission will fall onto the inner regions of the accretion disc. From these inner regions regions of the disc, photons pass closer to the black hole and so a greater fraction return to the disc. Fig.~\ref{return_fraction_h.fig} shows the escaping and returning fraction of the reflected photons as a function of the height of a point source above the singularity (which may be considered a proxy for the extent of the corona). The returning fraction increases from 35 per cent when the disc is illuminated by a source at $h=10$\rg\ to 47 per cent when the source is at $h=1.5$\rg. The total escape fraction by which the observed reflection fraction is reduced (Equation~\ref{escapefrac.equ}), shown by the solid line in Fig.~\ref{return_fraction_h.fig}, reduces from 57 per cent for $h=10$\rg\ to 39 per cent for $h=1.5$\rg.

X-ray reverberation timescales in Seyfert galaxies typically indicate distances between the corona and accretion disc between 2 and 9\rg\ \citep{kara_global}, hence, in the following analysis, we shall consider a simplified model of a point source at $h=5$\rg.

\section{The reprocessed spectrum}
\textsc{xillverRR} predicts the spectrum that is emitted from the accretion disc when it is illuminated by the returning primary reflection. It is based upon the \textsc{xillver} model of \citet{garcia+2010,garcia+2011,garcia+2013}, which solves the radiative transfer, energy balance and ionisation equilibrium of a Compton thick plasma to calculate the emergent reflection spectrum. While in \textsc{xillver}, the incident spectrum takes the form of a cut-off power law to predict the primary reflection spectrum, in \textsc{xillverRR}, the incident spectrum is the sum of the primary power law and the primary reflection spectrum described by \textsc{xillver}, and the reflection spectrum is computed for the total illumination.

\textsc{xillverRR} has two variable parameters; the energy shift of the returning primary reflection to account for Doppler shift and gravitational redshift between the primary and secondary reflections, and the flux ratio between the returning and primary continuum radiation. Illumination is assumed to be constant in time, thus the ionisation parameter, $\xi = 4\pi F / n$, is fixed (the ionisation is assumed to not change between the arrival of the primary and returning radiation). The reflection spectrum is also varied by the iron abundance in the disc, $A_\mathrm{Fe}$, relative to the Solar abundance, and the photon index, $\Gamma$, of the primary continuum.

In this study, we assume values in line with those measured in narrow line Seyfert 1 galaxies in which strong X-ray reflection, indicated by a relativistically broadened iron K$\alpha$ line, is seen from the inner regions of the accretion disc, such as 1H\,0707$-$495 or IRAS\,13224$-$3809. We take $\xi = 50$\ergcmps, $A_\mathrm{Fe} = 8$ and $\Gamma = 3$.

\begin{figure}
\centering
\includegraphics[width=85mm]{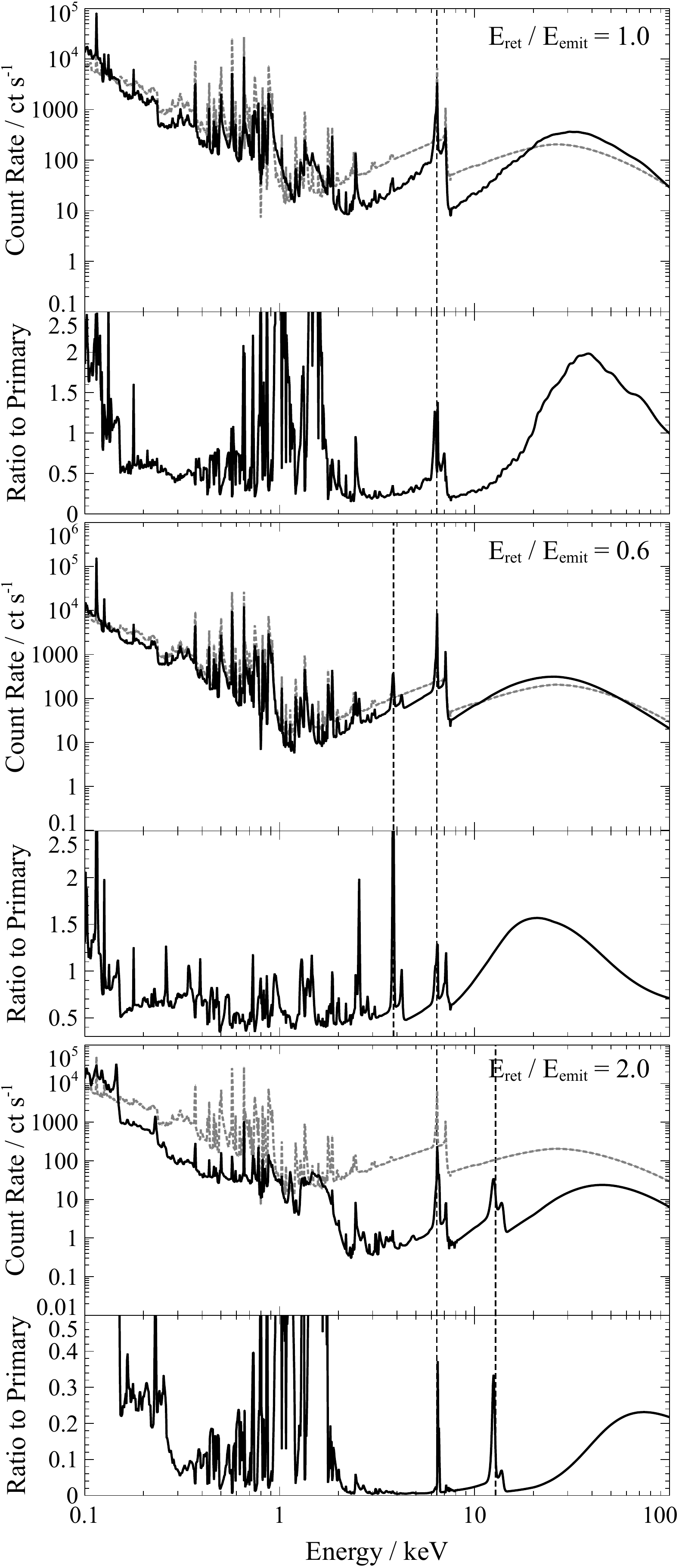}
\caption[]{The secondary reflection spectra, as seen in the rest frame of the reflecting material, for different energy shifts of the incoming primary reflection. $E_\mathrm{return} / E_\mathrm{emit} $ is the energy shift between the reflected photons from when they are emitted at the primary reflection site to when they are received after returning to the disc. In each case, the secondary reflection spectrum (solid curve) is compared to the primary reflection spectrum (dashed curve). Dashed vertical lines show the rest frame location of the 6.4\keV\ iron K$\alpha$ line and the position of the shifted 6.4\keV\ line in the incoming spectrum.}
\label{restframe_spec.fig}
\end{figure}

In order to illustrate the reprocessing of returning radiation by the accretion disc, Fig.~\ref{restframe_spec.fig} shows the secondary reflection spectrum as measured in the rest frame of emission, as predicted by \textsc{xillverRR}. The reprocessing of just the primary reflection is shown, with no incident continuum. Shown are the reprocessed spectra for three different energy shifts between the emission of the primary reflection spectrum and its return to the disc. In each case, the reprocessed spectrum is compared to the primary \textsc{xillver} reflection spectrum.

When there is no energy shift ($E_\mathrm{return} / E_\mathrm{emit} = 1$), the features of the reflection spectrum, notably the iron K$\alpha$ and $\beta$ lines around 6.4\keV\ and the Compton hump around 30\keV\ are sharpened, as discussed by \citet{ross_fabian_ballantyne}. The secondary reflection spectrum above 1\keV\ is suppressed, with greater relative suppression above the iron K absorption edge at 7.1\keV\ and less relative suppression of the core of the iron K$\alpha$ line. This results in the iron K absorption edge and fluorescence line being enhanced with respect to the rest of the spectrum, while sharpening the low energy side of the Compton hump.

Some of the most striking alterations to the spectrum occur when the returning radiation is blueshifted. Shown for illustration is an extreme case where $E_\mathrm{return} / E_\mathrm{emit} = 2$. In this case, the iron K line photons returning to the disc are shifted to 12.8\keV. At such high energies, there is little photoelectric absorption. As such, the majority of these blueshifted iron K line photons are simply scattered, producing a blueshifted iron K line at around 12\keV\ in the secondary reflection spectrum. This is slightly broadened as some of the scattered photons gain or lose energy by Compton scattering. This is produced in addition to the 6.4\keV\ fluorescence line that is produced after the photoelectric absorption of photons in the returning radiation field.

The returning radiation changes the effective ionisation balance in the upper layers of the disc in which the reflection spectrum is produced. The effective ionisation potential is increased by the increased incident flux from the returning radiation \citep{ross_fabian_ballantyne}. In the case of no energy shift, the slightly increased effective ionisation in the surface layers enhances the soft X-ray emission lines around 1\keV\ relative to the primary reflection spectrum.

When the returning radiation is blueshifted, the enhancement in the effective ionisation has a more significant effect that is particularly important with a steep, soft continuum spectrum (\textit{i.e.} when $\Gamma \gtrsim 2.5$). In this case, the hard X-ray spectrum is photon-starved, but blueshifting the primary reflection significantly increases the number of photons at high energies. This results in an increased temperature in the upper layers of the disc and a high effective ionisation paremeter. In addition to increasing the Compton broadening of the 12\keV\ blueshifted iron K line, this results in the smearing of the soft X-ray lines into a clearly defined soft excess.

When the returning primary radiation is redshifted, as in the case $E_\mathrm{return} / E_\mathrm{emit} = 0.6$, the alteration of the spectrum during the secondary reflection is more subtle. In this case, the redshift results in the hard X-ray spectrum being further photon starved, meaning the upper layers of the disc do not experience an increase in temperature and the effective ionisation potential is reduced. Therefore, when the returning radiation is strongly redshifted, the soft X-ray lines are not enhanced to the same degree. Some Compton scattering of the returning radiation does still occur, however, leading to  a redshifted reflection of the primary iron K lines around 3.8\keV\ in the secondary reflection spectrum, in addition to the soft X-ray lines in this part of the spectrum and the 6.4\keV\ iron K$\alpha$ fluorescence line. Similar behaviour is seen with more extreme redshifts of the returning radiation, though the redshifted, scattered iron K lines then sit among the soft lines and are not so intuitively identified as in this illustrative case.

\section{The observed spectrum}
The total observed spectrum was calculated from ray tracing simulations including both the primary illumination of the accretion disc by the corona and the illumination of the disc by the returning primary reflection. Higher order returning radiation would have added considerable computational expense and was not included in this study. From Equation~\ref{gp.equ}, each successive order is approximately 59 per cent weaker.

We consider the time-average spectrum, so therefore assume that the primary irradiation remains constant in time. The total reprocessed spectrum was computed for each radius on the disc for irradiation by the sum of the primary continuum and the returning primary reflection. For the emergent secondary reflection spectrum from each radial bin, we compute the average spectrum based upon the distribution of the energy shifts of the returning photons, weighted by the observed fluxes in the receiving frame of reference on the disc. Gravitational redshifts and Doppler shifts between each point on the disc and the observer producing the relativistic broadening have been applied.

\begin{figure}
\centering
\includegraphics[width=85mm]{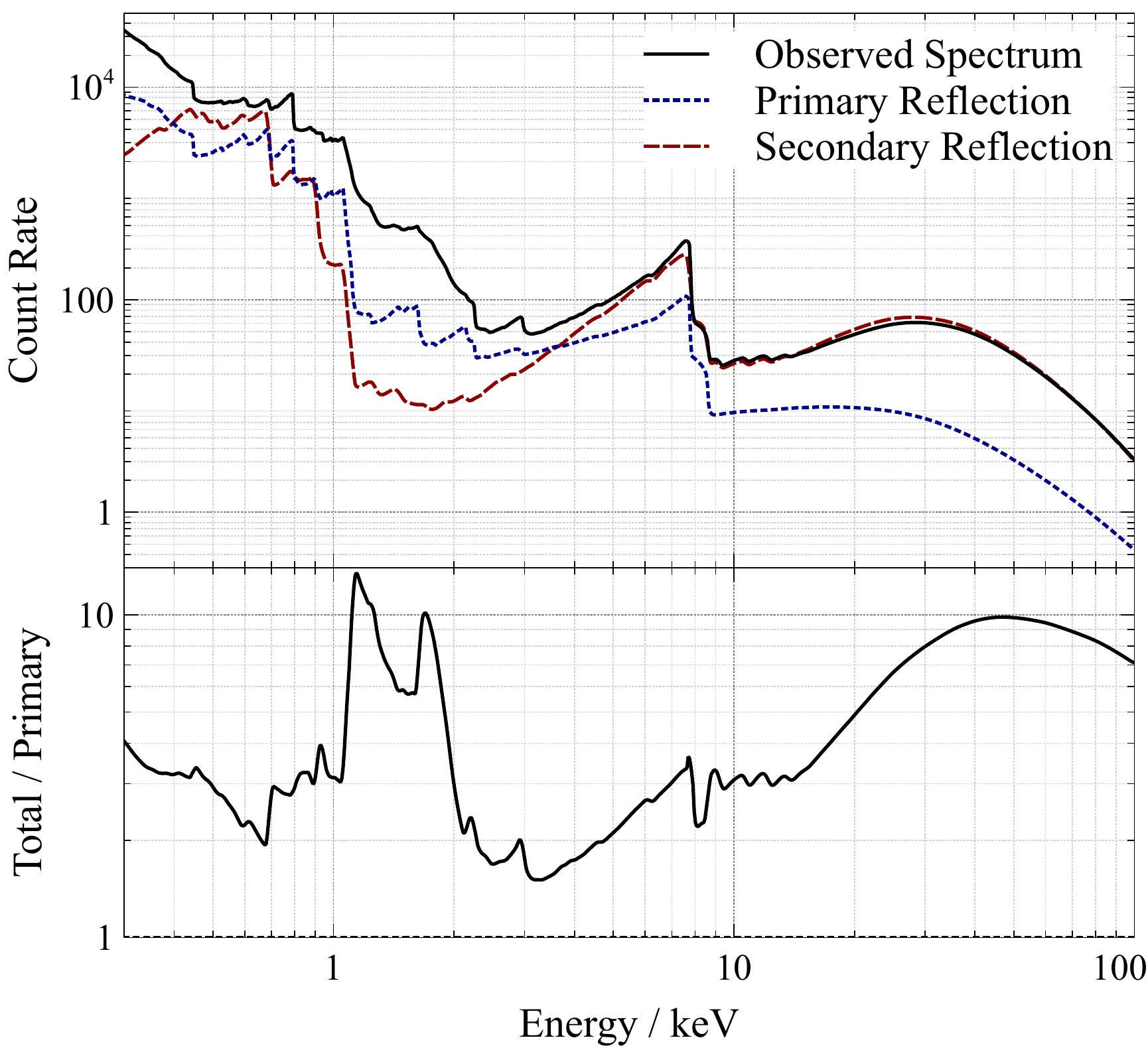}
\caption[]{The time-averaged spectrum of the primary and secondary reflection from the accretion disc, as observed from infinity, as well as the total observed reflection component. The lower panel shows the ratio of the total reflection spectrum to primary reflection. The spectra are shown for an accretion disc irradiated by a power law continuum spectrum with photon index $\Gamma = 3$. The iron abundance in the disc is eight times the Solar abundance, and the ionisation parameter $\xi = 50$\ergcmps. The disc is observed at an inclination of 60\,deg.}
\label{spectrum.fig}
\end{figure}

The observed spectrum is shown in Fig.~\ref{spectrum.fig} and is compared, for reference, to the relativistically broadened primary reflection spectrum, and to the total secondary reflection that would be produced if the disc were illuminated by only the returning primary reflection. The lower panel displays the ratio of the total observed reflection to just the primary reflection.

While returning radiation has a pronounced effect on the rest-frame reflection spectra emitted from individual annuli on the disc, the effects are much more subtle when the integrated spectrum from the whole disc is considered and the emission is subject to gravitational redshifts as it climbs out of the potential of the black hole. The most notable effect of returning radiation on the observed, time-averaged spectrum is an enhancement in the relativistically broadened iron K$\alpha$ line, slightly redward of the peak of the line. The strength of the line above the continuum is enhanced by 25 per cent. 

\begin{figure}
\centering
\includegraphics[width=85mm]{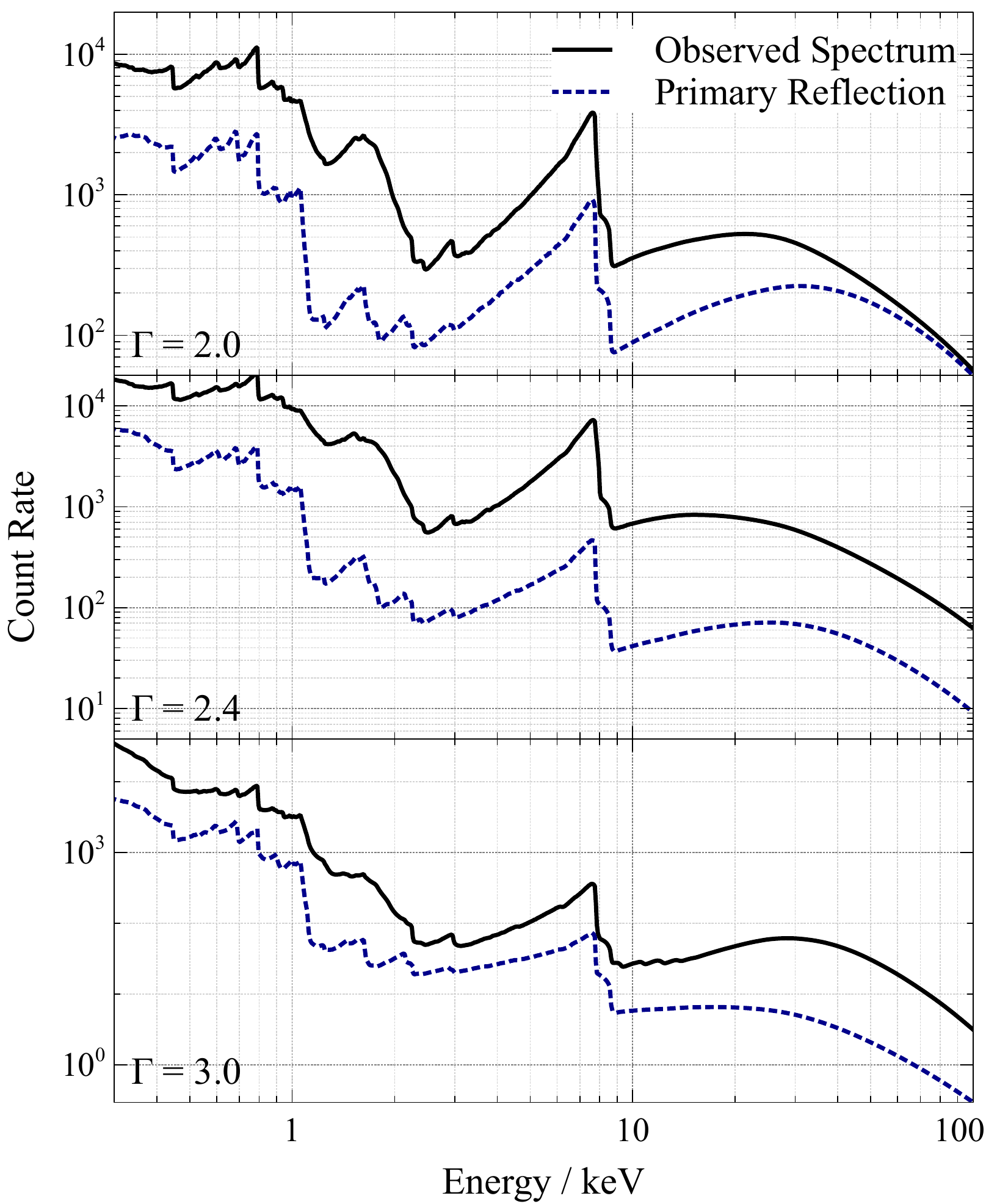}
\caption[]{The variation in the reflection spectrum as the photon index of the primary continuum, $\Gamma$, is varied. In each case, the total reflection spectrum, including both the primary reflection and the secondary reflection due to returning radiation is compared to the primary reflection spectrum that would be observed in the absence of returning radiation.}
\label{spectrum_gamma.fig}
\end{figure}

Much of the returning radiation is Compton scattered, enhancing the amplitude of the Compton hump (relative to the best-fitting underlying continuum) by a factor of three. This is particularly pronounced when the accretion disc is illuminated by a steep continuum spectrum, as in the case of the narrow line Seyfert 1 galaxies in which the strongest reflection is seen. When the photon index $\Gamma = 3$, as shown here, the hard X-ray spectrum is photon-starved and few photons are available at high energy to Compton scatter. The blueshifting of the returning radiation to higher energy significantly increases the population of photons that form the Compton hump. When the continuum spectrum is harder, the enhancement of the Compton hump by returning radiation is less. For $\Gamma = 2.4$, the level of the Compton hump above the best-fitting power law is enhanced by a factor of two, and only below 30\keV, as shown in Fig.~\ref{spectrum_gamma.fig} (the high energy part of the Compton hump is not enhanced when the primary X-ray continuum is harder). For $\Gamma = 2$, the Compton hump is enhanced by only a factor of just 1.35 below 20\keV\ and is, in fact, slightly weakened at high energy, which may mimic a lower coronal temperature. In each case, the iron K absorption edge is deepened, sharpening the low energy side of the Compton hump.

Because a significant fraction of the returning radiation is blueshifted, the ionisation balance is changed in the layers of the disc in which the reflection spectrum is produced. The increase in the effective ionisation produces additional soft X-ray emission, particularly in the 1-2\keV\ band. Such additional soft X-ray emission has been seen in the spectra of a number of AGN \citep{jiang_iras} and has often been modeled by including additional reflection components (attributed to a clumpy structure in the disc with varying ionisation parameter across the surface) or by an increase in the disc density. Returning radiation may explain some of this excess emission and reduce the need for such a severe increase in disc density.

\begin{figure}
\centering
\includegraphics[width=85mm]{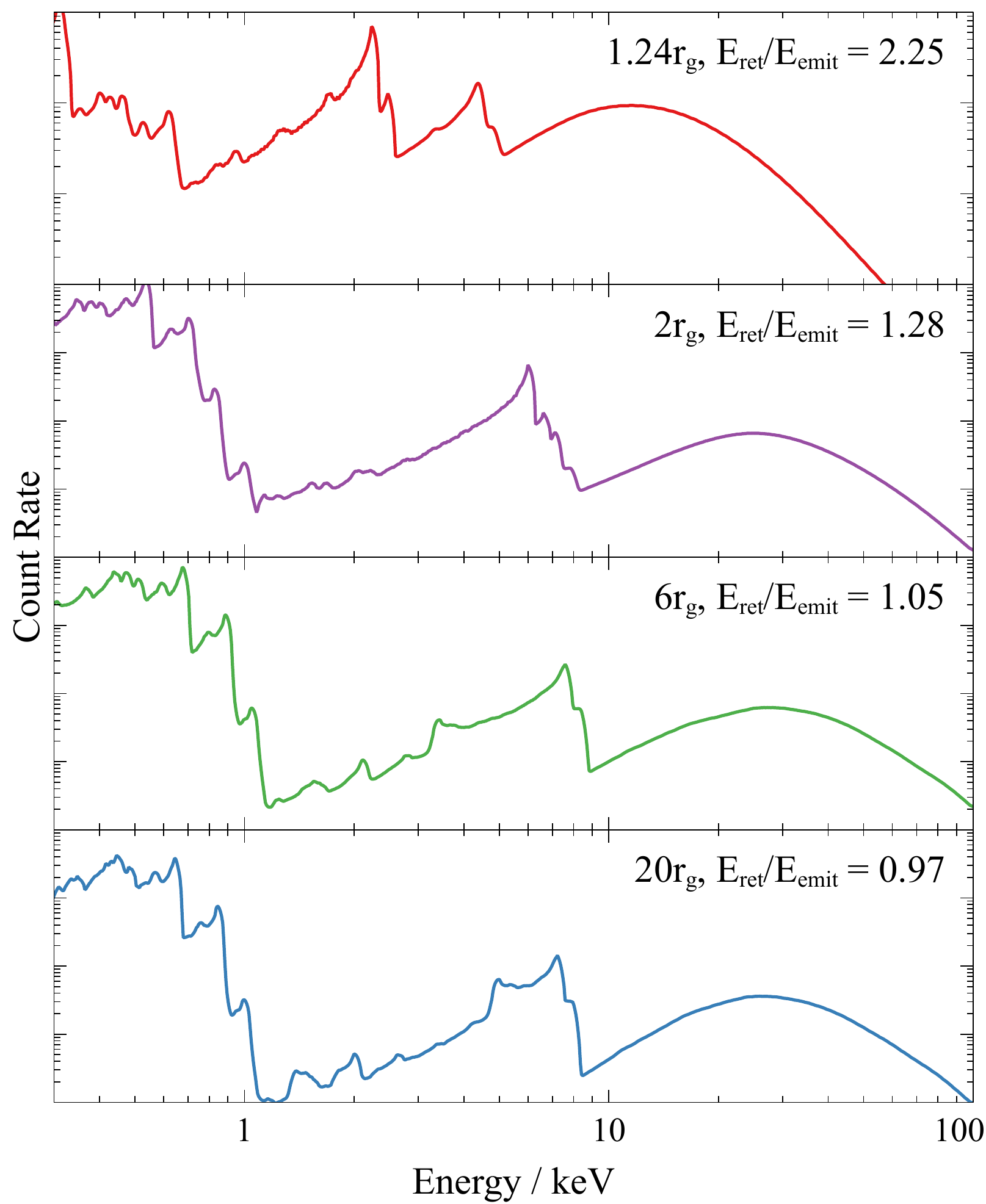}
\caption[]{The secondary reflection spectra emitted from annuli between 1.24 and 2\rg\ as observed at infinity, including the observed Doppler shifts and gravitational redshift. For each annulus, the average energy shift of the returning primary reflection is shown.}
\label{return_radspec.fig}
\end{figure}

The secondary reflection spectrum that is \textit{observed} from different radii, after the effects of the Doppler shift and gravitational redshift between the disc and the observer, is shown in Fig.~\ref{return_radspec.fig}. Even though a strong blueshifted iron K line is produced when the returning radiation is blueshifted, the returning radiation most strongly blueshifted only within the innermost 2\rg\ around a maximally spinning black hole. The photons emitted from such small radii are all strongly redshifted as they climb out of the gravitational potential, which means that the blueshifted iron K line is seen not at 12\keV\ but between 2.5 and 5\keV. This broadened, blueshifted, Compton-scattered returning line can be seen separated from the iron K$\alpha$ fluorescence linein the emission from 1.24\rg\ on the disc, while at 2\rg\ the blueshifted line blends into the blue wing of the fluorescence line. At 6\rg, the returning radiation is no longer blueshifted and the additional scattered line emission can no longer be seen. A subtle peak corresponding to the scattered, blueshifted, returning iron K photons can be seen in the total reflection spectrum, forming a peak between 2.5 and 3.4\keV, though the peak is only 30 per cent above the emission features that surround it, which means it is likely difficult to detect in the time-averaged spectrum.

The shape of the reflection spectrum depends upon the inclination at which the accretion disc is viewed. Most notably, increasing the inclination angle of the disc normal to the line of sight (\textit{i.e.} viewing the disc closer to edge-on) increases the range of radial velocities that are observed which hence broadens the profile of the iron K$\alpha$ line. The enhancements in the reflection spectrum due to returning radiation are not strongly dependent on the inclination angle. Decreasing the inclination angle from the pictured 60\,deg to 30\,deg produces a soft excess, an iron K line and a Compton hump that are enhanced to approximately the same degree.

\section{X-ray reverberation}
While the signatures of reflected X-ray emission returning to the accretion disc through strong light bending may be difficult to detect directly in the time-averaged spectrum, the signatures may be manifested more clearly in the time delays between different energy bands, in the X-ray reverberation signals that are seen from the inner accretion disc.

The X-ray continuum emitted from the corona is highly variable. Variability in the continuum luminosity will be mirrored in the intensity of reflection signatures after a time lag corresponding to the additional light travel time between the corona and the accretion disc \citep{reverb_review}. Signatures of returning radiation will be further delayed due to the additional light travel time between the sites of the primary and secondary reflection.

Reverberation signatures of returning radiation were explored by considering the response of the accretion disc to a single flash of X-rays emitted from the corona. In this case, the primary illumination will arrive at each point on the disc before any returning radiation, hence the reprocessed spectrum in response to each can be considered separately, rather than having to account for a time-variable primary to returning flux ratio.

When the accretion disc is irradiated by the continuum, species will become ionised and the reprocessed spectrum will vary accordingly on the order of the ionisation timescale, $t_\mathrm{ion} \sim \left<h\nu\right> / F_X \sigma_\mathrm{ion} \sim 10^{-7} (F / 10^{16}\mathrm{erg\,cm^{-2} s^{-1}})^{-1}\,\mathrm{s}$ \citep{collin+2003}. Approximating Euclidean geometry, the flux received at radius $r$ on the disc is $L_X / 4\pi(r^2 + h^2)$. Thus, assuming a $10^7$\Msun\ black hole about which the X-ray luminosity is 2 per cent of the Eddington luminosity (about 10 per cent of the bolometric luminosity where $L_\mathrm{bol}\sim 0.2$ to maintain a thin accretion disc), \textit{i.e.} $L_X\sim 0.02L_\mathrm{Edd}$, the ionisation timescale at $r=10$\rg\ in a disc illuminated by a point source at $h=5$\rg\ is $t_\mathrm{ion} \sim 10^{-7}\,\mathrm{s}$.

Following the flash of continuum emission, ions and electrons in the disc will recombine and the reprocessed emission from the disc will `reset' to its initial state before the flash on the recombination timescale; $t_\mathrm{rec}\sim 1/n\alpha_\mathrm{rec} \sim 10(n / 10^{12}\mathrm{cm^{-3}})^{-1}\,\mathrm{s}$. When the number density of the plasma in the accretion disc is taken to be $n = 10^{15}\,\mathrm{cm}^{-3}$, as is typically assumed in reflection models applicable the accretion discs around supermassive black holes \citep[\textit{e.g.}][]{garcia+2013}, the recombination time $t_\mathrm{rec}\sim 0.01$\s. If the density is increased to $n = 10^{19}\,\mathrm{cm}^{-3}$, as suggested by recent spectral analysis of the reflected X-rays from the accretion disc in the narrow line Seyfert 1 galaxy IRAS\,13224$-$3809 \citep{jiang_iras}, the recombination time shortens to $t_\mathrm{rec}\sim 10^{-6}$\s.

The interval between the arrival of the primary continuum and the mean arrival time of the returning radiation (from all parts of the disc) is minimum on the inner disc. Here, the delay between the mean arrival time of the primary and secondary reflection is $22\,GMc^{-3}$ where the natural time unit $GMc^{-3}$ is the light travel time over one gravitational radius, and can be readily scaled by the mass of the black hole to evaluate light travel times for any black hole mass. For a $10^{7}$\Msun\ black hole, $GMc^{-3} = 50$\s, much longer than either the ionisation or the recombination timescale. The disc can be assumed to have reset between the arrival of the primary continuum and the returning radiation and these can be treated separately in the simulation when just a single flash of emission from the corona is considered.

\subsection{The response of the disc}

\begin{figure}
\centering
\subfigure[Primary response] {
\includegraphics[width=85mm]{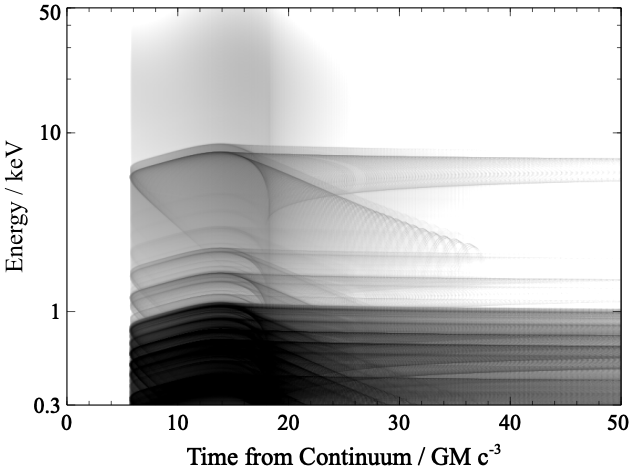}
\label{ent.fig:primary}
}
\subfigure[Secondary response (returning radiation)] {
\includegraphics[width=85mm]{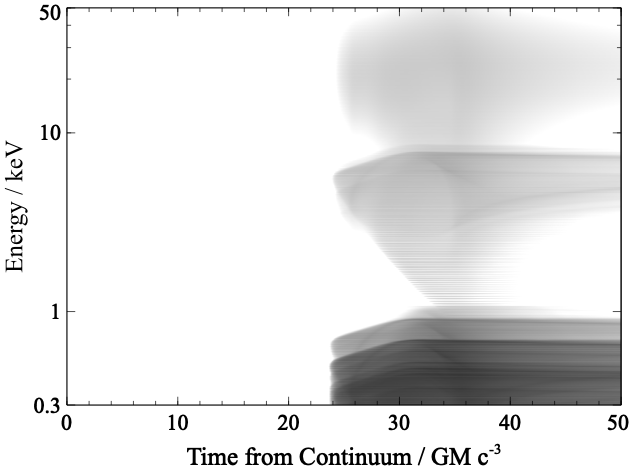}
\label{ent.fig:return}
}
\subfigure[Total response] {
\includegraphics[width=85mm]{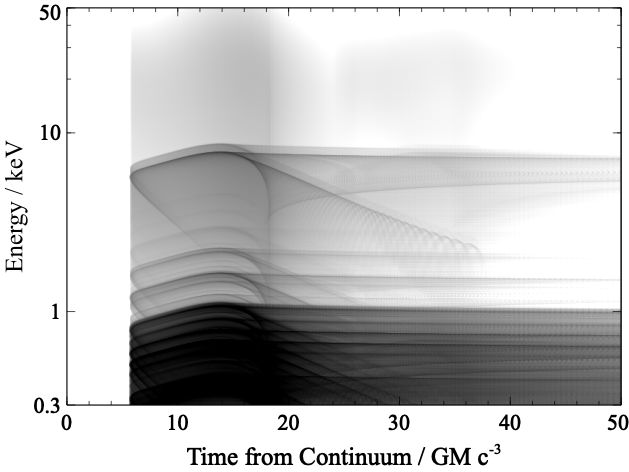}
\label{ent.fig:total}
}
\caption[]{The time and energy resolved impulse response function of the accretion disc, reverberating in response to a single flash of emission from a coronal point source located 5\rg\ above the singularity. \subref{ent.fig:primary} shows the primary reverberation component, \subref{ent.fig:return} the secondary component and \subref{ent.fig:total} the total response. The primary continuum flash is seen from the corona at $t=0$.}
\label{ent.fig}
\end{figure}

Fig.~\ref{ent.fig} shows the energy- and time-resolved response functions of the reverberation seen from the accretion disc after a single flash of X-rays from the corona. The total response is shown alongside the primary and secondary responses. Though each shows the characteristic response of relativistically broadened emission lines; the iron K lines around 6.4\keV\ and the lines that comprise the soft X-ray emission, discussed in detail by \citet{reynolds+99} and \citet{propagating_lag_paper}, the additional time delay between the arrival of the majority of the primary and the secondary emission is immediately apparent.

Following the flash of continuum emission, a single response is seen in each emission line in the primary reverberation component. The earliest arriving line photons are seen slightly redshifted from the rest frame energy, originating just beyond the inner parts of the disc (around 5\rg) where the integrated light path from the corona to the disc to the observer is minimised, including the effects of light travel time delays through the strong gravitational field. The most red- and blueshifted line emission comes from the innermost parts of the disc, where the orbital velocity is the greatest. Light travel time delays mean the response function broadens to reach the most extreme red- and blueshifts some time after the first response is seen. The blueshifted side of the line response is seen to loop round to low, redshifted energies as gravitationally lensed (hence magnified) emission from the back side of the disc re-emerges from behind the black hole. Finally, the latest response is seen in a double-peaked line profile from the approaching and receding sides of the outer disc, furthest from the primary X-ray source. The reflected continuum provides emission over all energies over a range in time, while the Compton hump produces broad, delayed emission feature around 30\keV.

While the primary reverberation shows just a single response in each emission line, the returning radiation produces multiple line responses superimposed upon one another at later times, becoming broader in their range of redshift. In addition, the returning, blueshifted iron K photons that are scattered from the innermost parts of the disc add an additional line-like feature starting at 3\keV\ when it is first observed, that is not seen in the primary response. Like an emission line response, this broadens in energy at later times, but arises only from the innermost part of the disc, hence no corresponding late-time narrow response is seen.

\subsection{Detecting the reverberation of returning radiation}

\subsubsection{The lag-energy spectrum}

\begin{figure}
\centering
\includegraphics[width=85mm]{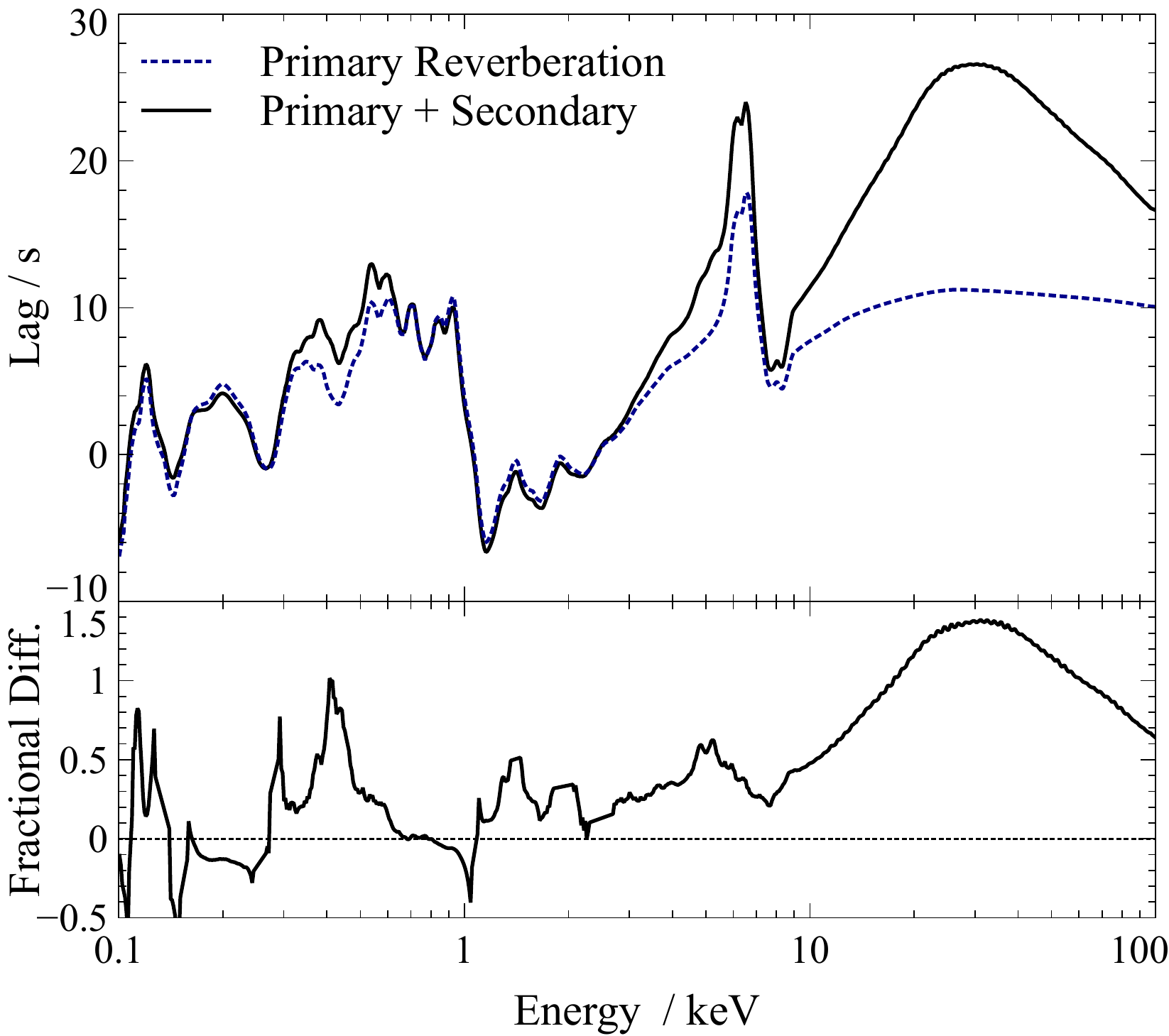}
\caption[]{The lag-energy spectrum showing the average observed response time of the accretion disc as a function of X-ray energy. The summed primary and secondary response is compared to just the primary response.}
\label{avg_arrival.fig}
\end{figure}

Fig.~\ref{avg_arrival.fig} shows the average response time as a function of energy, relative to the response time in the most continuum-dominated band between 1 and 2\keV. This is the equivalent of the \textit{lag-energy spectrum} from which reverberation measurements are made in X-ray observations (though the lag-energy spectrum shows the time lag over a range of Fourier frequencies, whereas the average response time would be the lag averaged across \textit{all} Fourier frequencies if reverberation were the only process imparting lags between energy bands). The response of the total continuum, primary and secondary reverberation is compared to just that of the continuum plus primary reverberation.

For X-rays originating from a point source at $h=5$\rg\ above the singularity, the inclusion of the secondary response from returning radiation increases the reverberation time lag across the broad iron K line, due to the additional iron K line emission from returning radiation arriving at late times. The average arrival time of photons in the core of the iron K line at 6.4\keV\  is increased by just under 30 per cent, while the average arrival time of photons in the redshifted wing is increased by 60 per cent. This increase in the arrival time of the red wing of the line relative to the core of the line distinguishes the change in the lag profile from that caused by variation in either the black hole mass or the scale height of the primary X-ray source which would cause a rescaling of the entire iron K line lag profile \citep{cackett_ngc4151}.

Between 0.5 and 1\keV, the soft X-ray reverberation lag is, on average, not affected by the presence of returning radiation. This is not to say that the secondary soft X-ray reflection is not delayed, rather that the lag is shorter relative to the 1-2\keV\ band. When returning radiation is included, additional reprocessed emission between 1 and 2\keV\ from the ionised top layers of the disc delays the average response time of this reference band.

The response time of the Compton hump is more than doubled at energies between 10 and 30\keV\ as the returning radiation contributes significantly to this band. The strong iron K absorption edge in the secondary reflection spectrum, however, means that emission between 7 and 9\keV\ is strongly dominated by the continuum and primary reverberation components (a signiificant number of these photons are absorbed in the secondary reverberation), producing a sharp dip in the lag profile between these energies, between the iron K emission line and the rise of the response time into the Compton hump.

The scattered blueshifted iron K line photons that return to the inner disc are manifested in the lag-energy spectrum as a steepening in the rise of the iron K line between 3.5 and 4.5\keV. These photons are scattered within the innermost 2\rg\ of the disc. They are redshifted before they reach the observer, and their passage is significantly delayed as they travel through the strong gravitational field \citep{shapiro}.

\subsubsection{Time domain response functions}

\begin{figure*}
\centering
\subfigure[0.3-1\keV (soft excess)] {
\includegraphics[width=78mm]{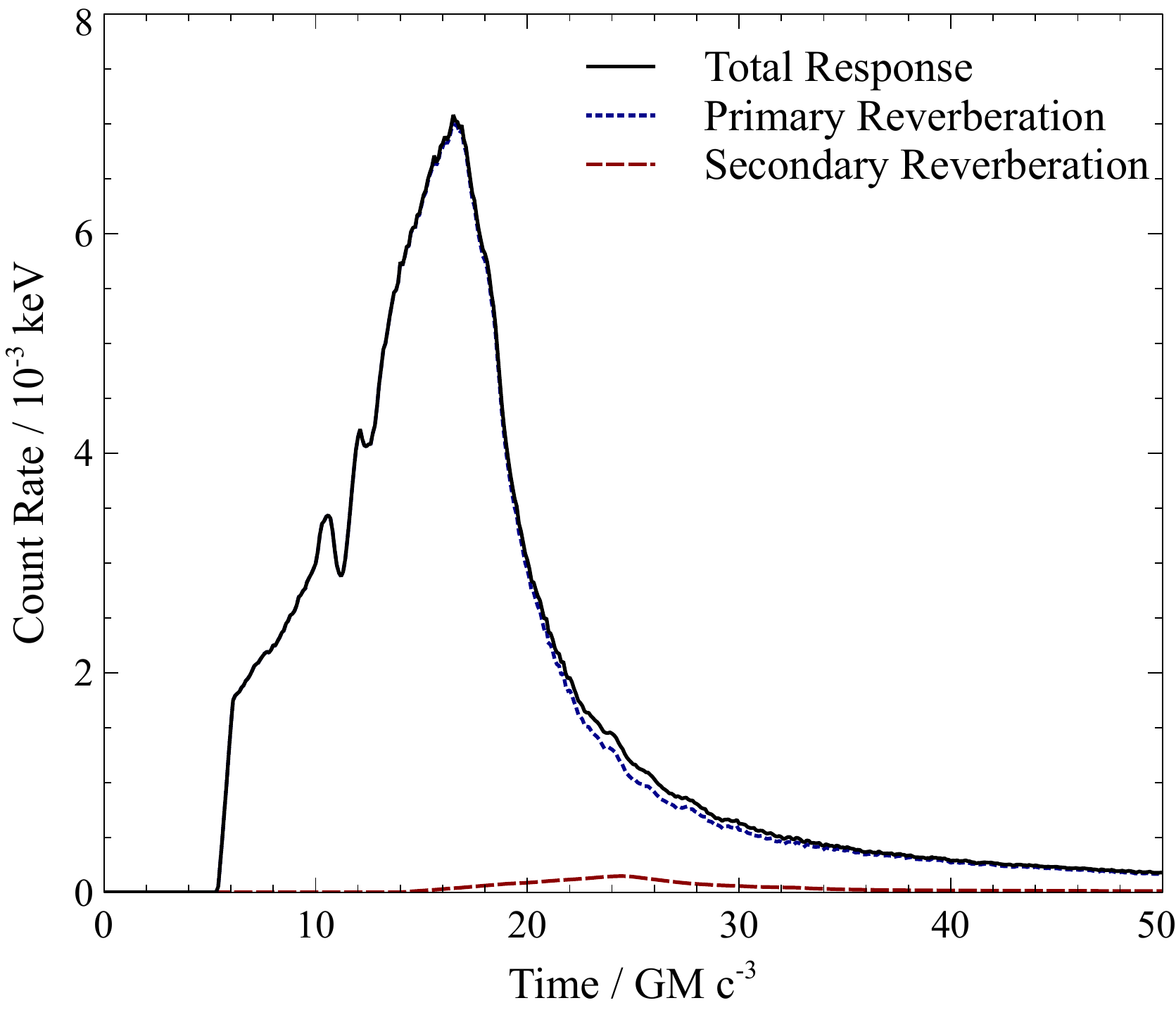}
\label{resp.fig:0.3-1}
}
\subfigure[5-7\keV (iron K line)] {
\includegraphics[width=78mm]{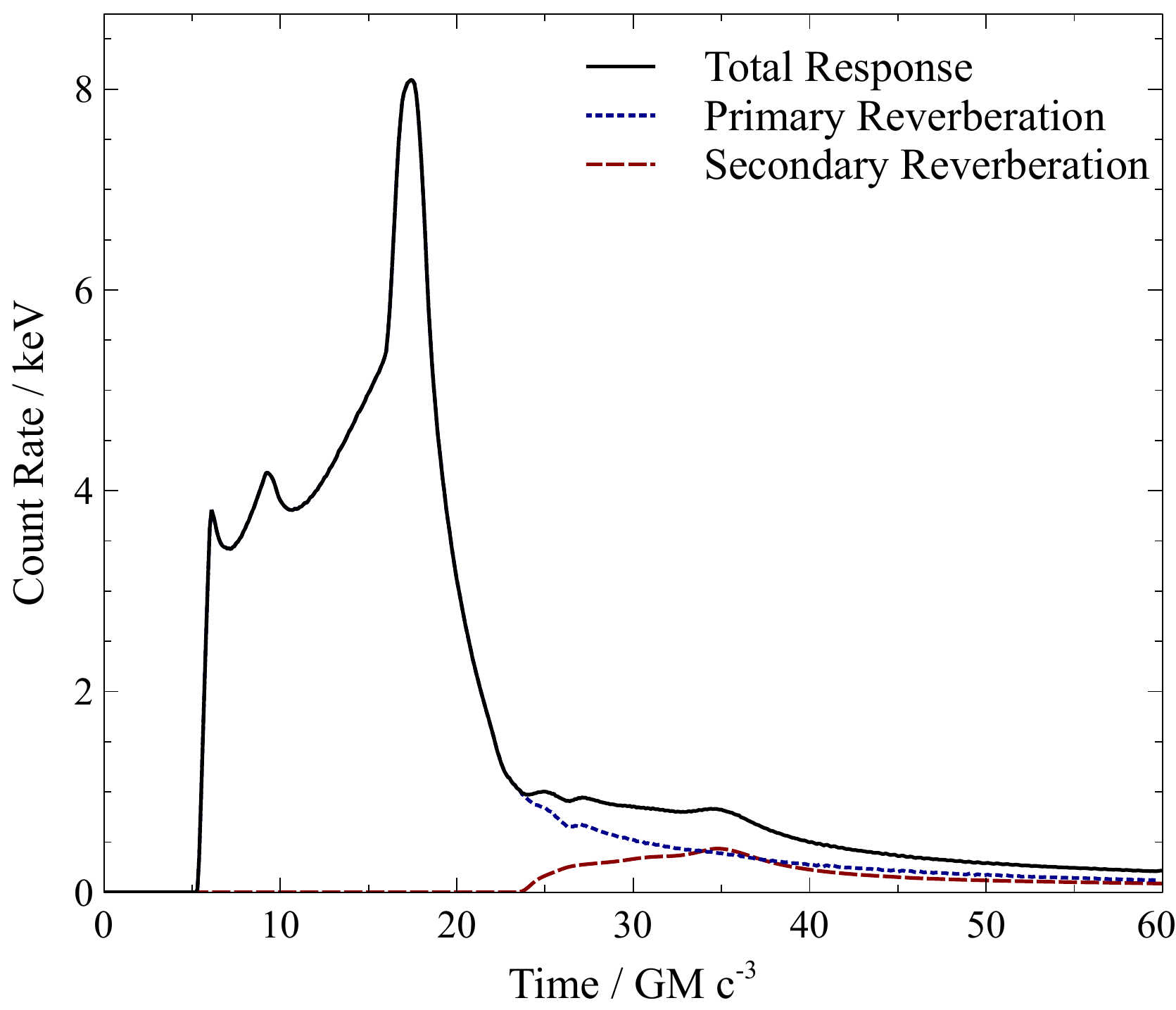}
\label{resp.fig:5-7}
}
\subfigure[10-30\keV (Compton hump)] {
\includegraphics[width=78mm]{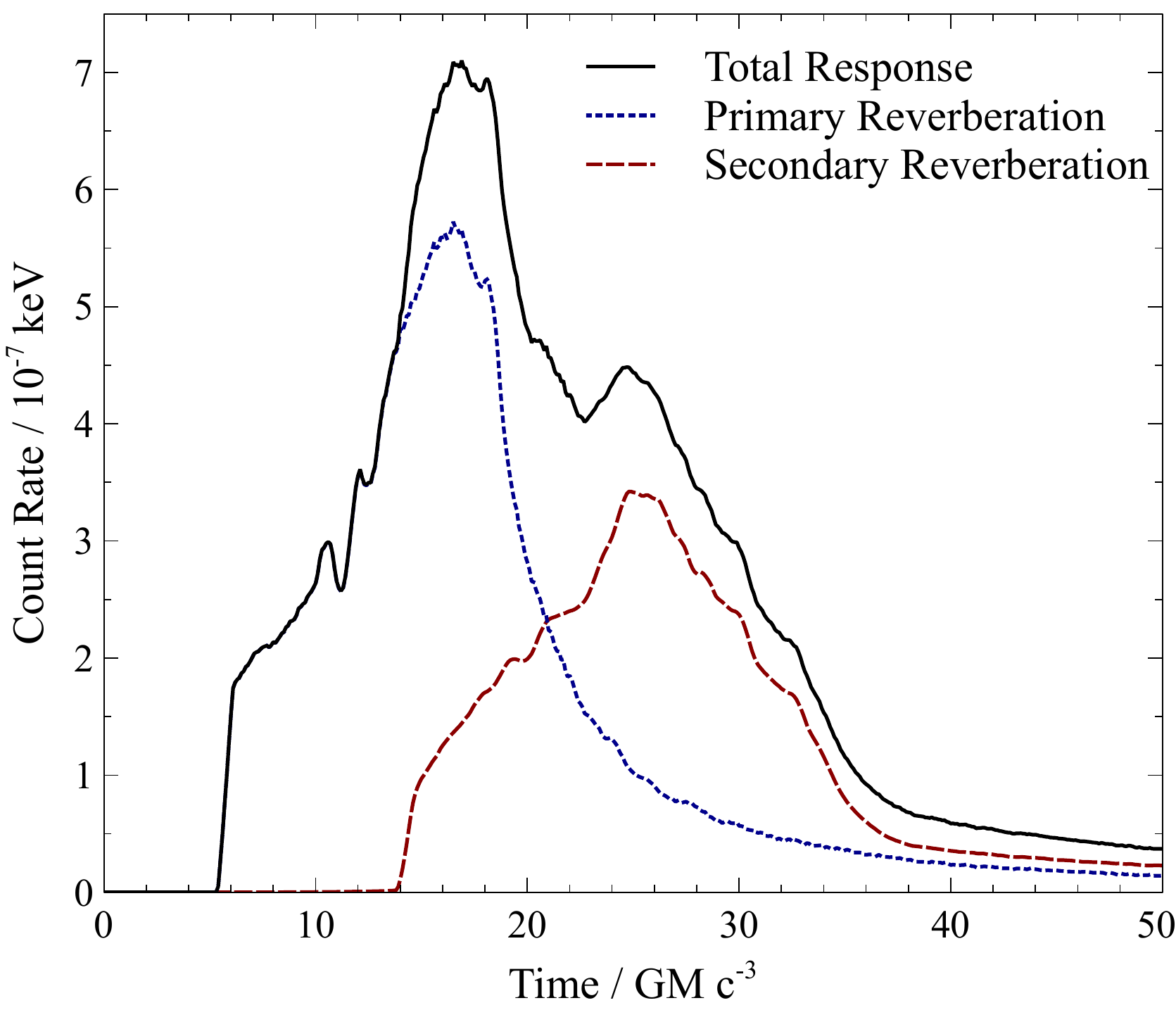}
\label{resp.fig:10-30}
}
\subfigure[3.5-4.5\keV (returning blueshifted Fe K)] {
\includegraphics[width=78mm]{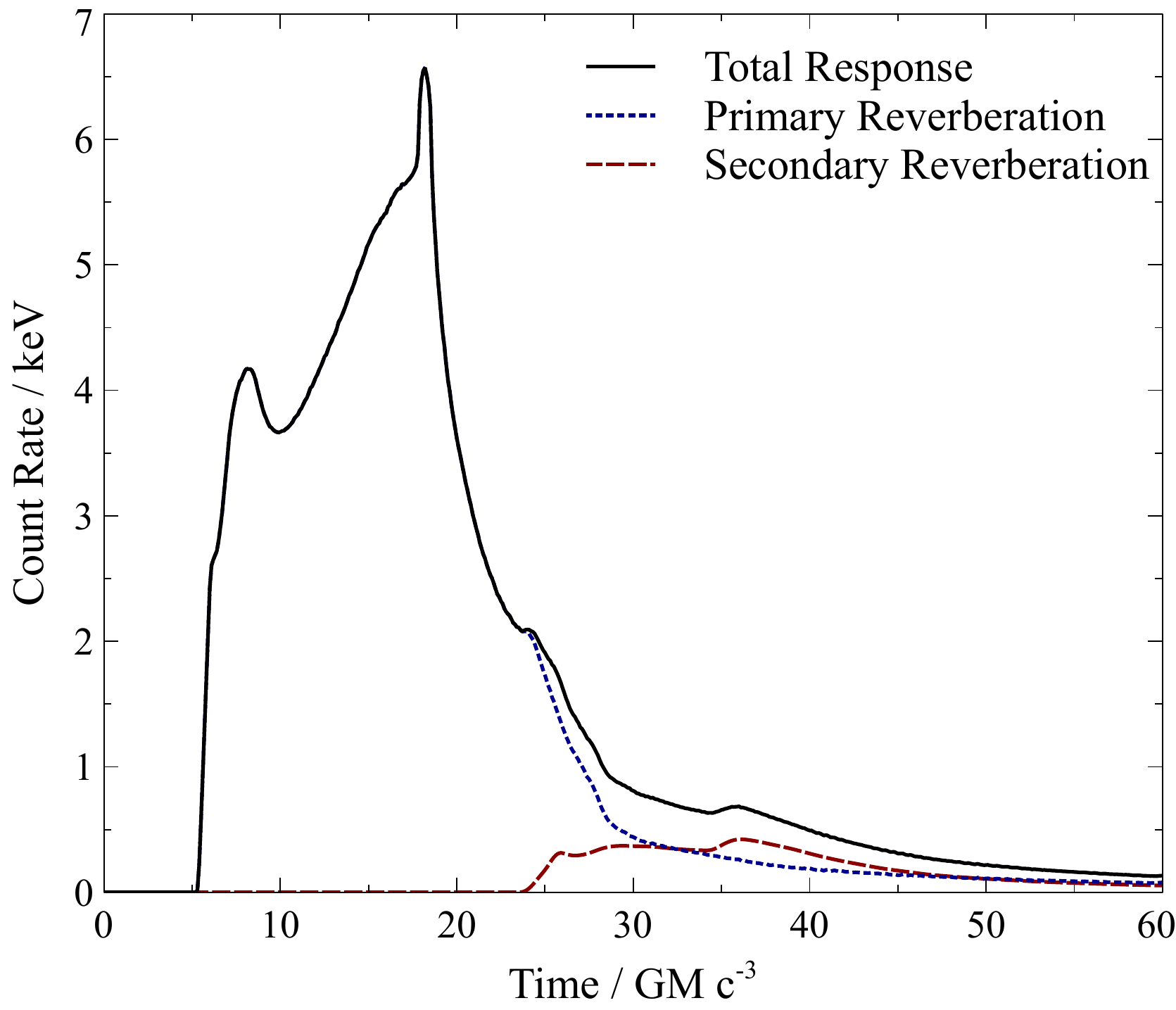}
\label{resp.fig:3.5-4.5}
}
\caption[]{The impulse response function of the accretion disc in response to a single flash of emission from a point source at $h=5$\rg. as a function of time after the primary continuum emission is observed in energy bands corresponding to \subref{resp.fig:0.3-1} the soft excess emission between 0.3 and 1\keV, \subref{resp.fig:5-7} the core of the broadened iron K line between 5 and 7\keV, \subref{resp.fig:10-30} the Compton hump, and \subref{resp.fig:3.5-4.5} the 3.5-4.5\keV\ energy band containing redshifted iron K line emission and scattered, blueshifted iron K emission returning to the inner disc that is observed redshifted. In each band. the primary and secondary responses are shown in addition to the total response. The response functions are normalised according to the width of each band and the normalisation is consistent between each panel.}
\label{resp.fig}
\end{figure*}

In order to verify that any additional lag seen in the redshifted wing of the iron K line is due to the reverberation of returning radiation, rather than differences in the primary spectrum or the geometry of the corona or accretion disc that may also change the line profile, we examine the specific shape of the response in this energy band. Fig.~\ref{resp.fig} shows the time domain response functions to a flash of continuum emission at $t=0$, showing the total response and the contributions from the primary and secondary reverberation. The response is shown in energy bands corresponding to the soft excess, below 1\keV, the continuum, the core of the iron K line, and the blueshifted returning iron K photons scattered from the inner disc, superimposed upon the redshifted wing of the line.

The delayed response due to returning radiation can be clearly seen in energy bands dominated by the reflected iron K line complex, in Fig.~\ref{resp.fig:5-7} and \ref{resp.fig:3.5-4.5}. The primary response of the 5-7\keV\ iron K band initially peaks at time $6\,GMc^{-3}$ after the flash of primary continuum emission from a source at $h=5$\rg\ is received and is relatively narrow. Following this, there is a sharp second peak, consisting of the magnified, lensed emission re-emerging from the back side of the disc, which would classically be hidden behind the black hole. In strong gravity, these rays are bent around the black hole \citep{reynolds+99,lag_spectra_paper,cackett_ngc4151}. Following this re-emergence peak, the response decays at late times as the parts of the disc furthest from the primary source respond.

While the first primary response is seen at $t = 6\,GMc^{-3}$ after the continuum flash, the secondary reverberation is not seen until $t = 23\,GMc^{-3}$, creating a secondary peak at $t = 35\,GMc^{-3}$, reaching 25 per cent of the first part of the primary response (not including the sharp narrow peak of X-rays re-emerging from the back side of the disc). The secondary response is broad due to the multiple light paths by which reflected emission can return to/from across the disc. 

The width of the response may be quantified by the `half-flux interval,' the interquartile time interval between the times at which 25 and 75 per cent of the total cumulative flux have been received. The half-flux interval of the primary iron K response in the 5-7\keV\ band  is $9\,GMc^{-3}$ when the point source is located at $h=5$\rg. The inclusion of returning radiation increases the width of the response function to a half-flux interval of $15\,GMc^{-3}$.

The blueshifted iron K line photons returning to the inner regions of the disc and scattered before being redshifted on their passage to the observer are seen in the 3.5-4.5\keV\ band. These scattered returning photons form a broad secondary response that extends the decaying tail of the primary response (in this energy band, the primary response consists of the most strongly redshifted line photons from the inner disc in addition to a contribution from the reflected continuum). The half-flux interval is increased from $9\,GMc^{-3}$ to $12\,GMc^{-3}$, increased by slightly less than the width of the response in the 5-7\keV\ band since a narrower range of ray paths are able to produce this blueshifted emission returning to the inner disc, though the relative flux fraction is greater since the primary reflected flux in this energy band is lower.

Returning radiation has a much smaller impact on the soft excess, shown in Fig.~\ref{resp.fig:0.3-1}, where the half-flux interval is increased from $8\,GMc^{-3}$ to just $10\,GMc^{-3}$ and the secondary response peaks at just 2 per cent of the primary peak. On the other hand the effect on the low energy side of the Compton hump, shown in Fig.~\ref{resp.fig:10-30}, is much greater, with not just the average response time increasing by a factor of 2, but the half-flux interval increasing from $8\,GMc^{-3}$ to $21\,GMc^{-3}$, though with a strong secondary peak comprising the secondary reverberation. In the absence of returning radiation, the Compton hump responds 35 per cent sooner than the core of the iron K line, but when returning radiation is included, the response times of the core of the iron K line and of the iron K line core are comparable. The effect on the continuum-dominated 1-2\keV\ band is minimal and the half flux interval in this band is unchanged.

\subsubsection{Lag-frequency spectra}
The time domain impulse response functions represent the count rate as a function of time following a single flash of emission from the corona. Rather than observing a single flash of continuum emission, the accretion disc is observed responding to stochastic variability in the X-ray continuum. In this case, the observed light curve in each energy band is given by the time series describing the variability in the primary continuum convolved with the response in that energy band.

In X-ray observations, the response of the disc to stochastic variability in the continuum is measured in the Fourier domain. The response function is encoded in the \textit{cross-spectrum} between the Fourier transform of the light curve in the energy band of interest and of a reference band (in fact the response \textit{functions} of both energy bands are encoded). This is expressed as the \textit{lag-frequency} spectrum; the phase of the complex cross-spectrum, \textit{i.e.} the time lag as a function of the low and high Fourier frequency components that make up the observed variability \citep{zoghbi+09,reverb_review}.

\begin{figure*}
\centering
\subfigure[Soft Lag, 1-4\keV\ \textit{vs.} 0.3-1\keV] {
\includegraphics[width=55mm]{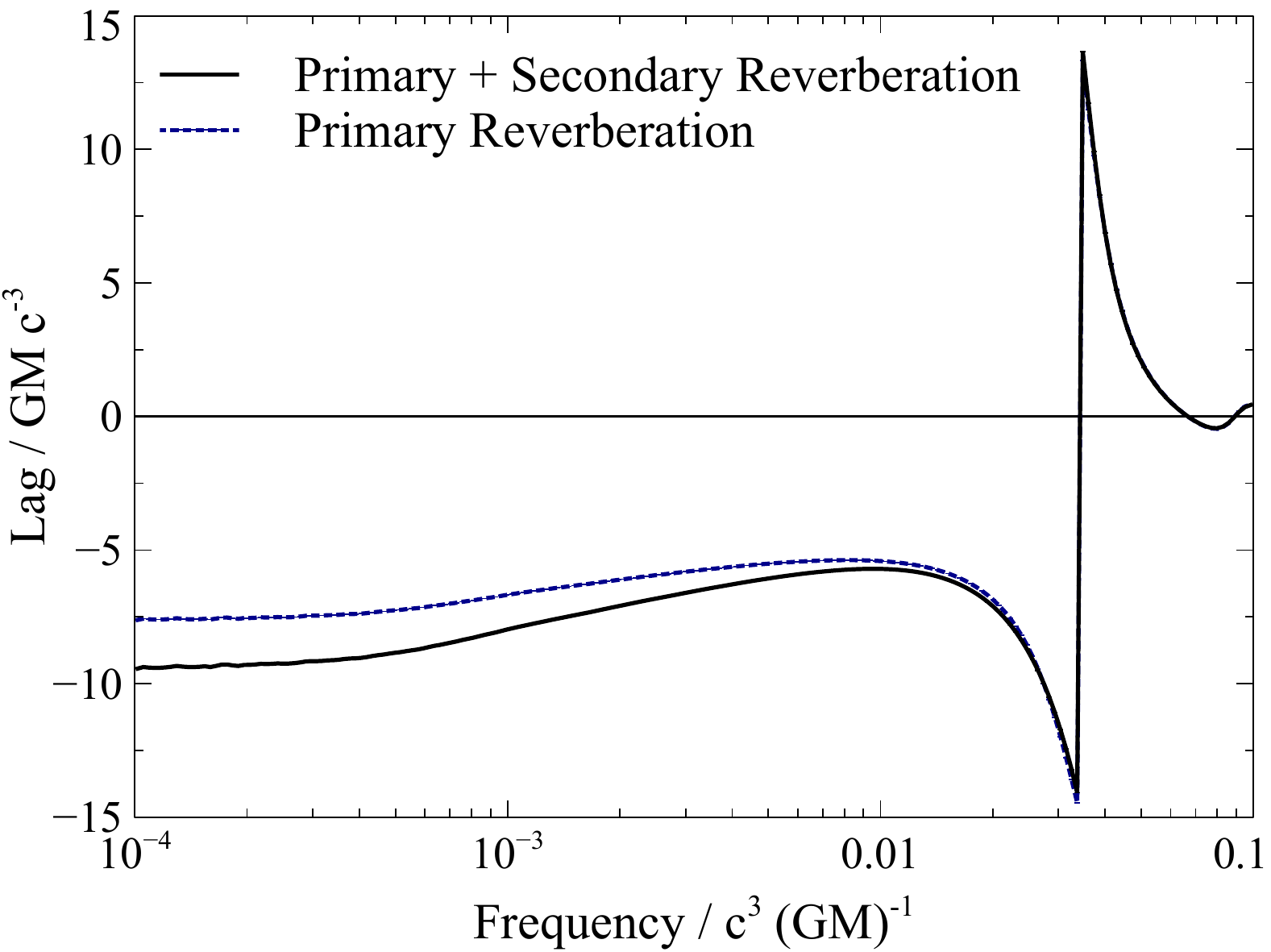}
\label{lagfreq.fig:soft}
}
\subfigure[Fe K lag, 4-7\keV\ \textit{vs.} 1-4\keV] {
\includegraphics[width=55mm]{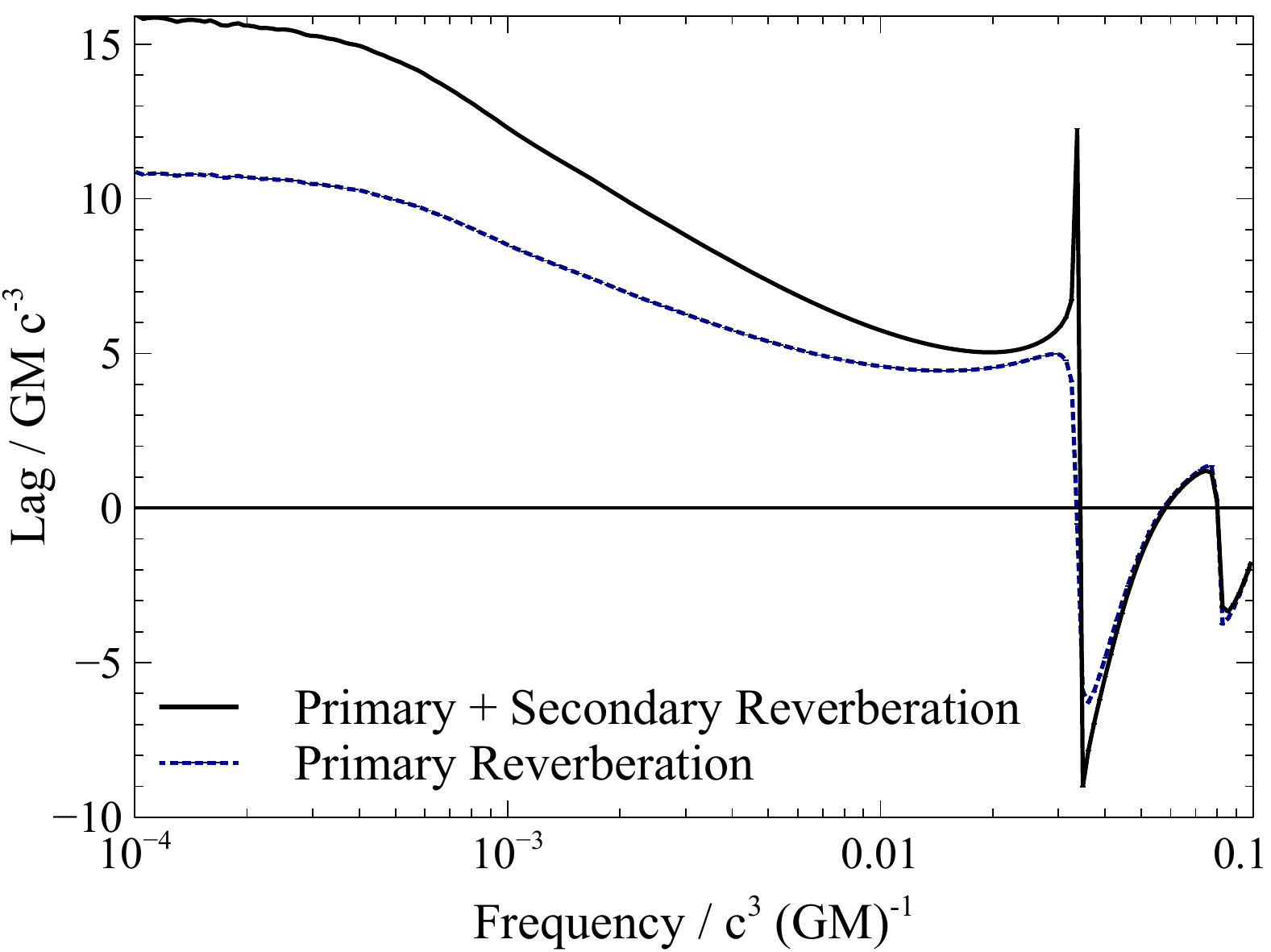}
\label{lagfreq.fig:fek}
}
\subfigure[Returning blueshifted Fe K, 3.5-4.5\keV\ \textit{vs.} 5-7\keV\ Fe K line] {
\includegraphics[width=55mm]{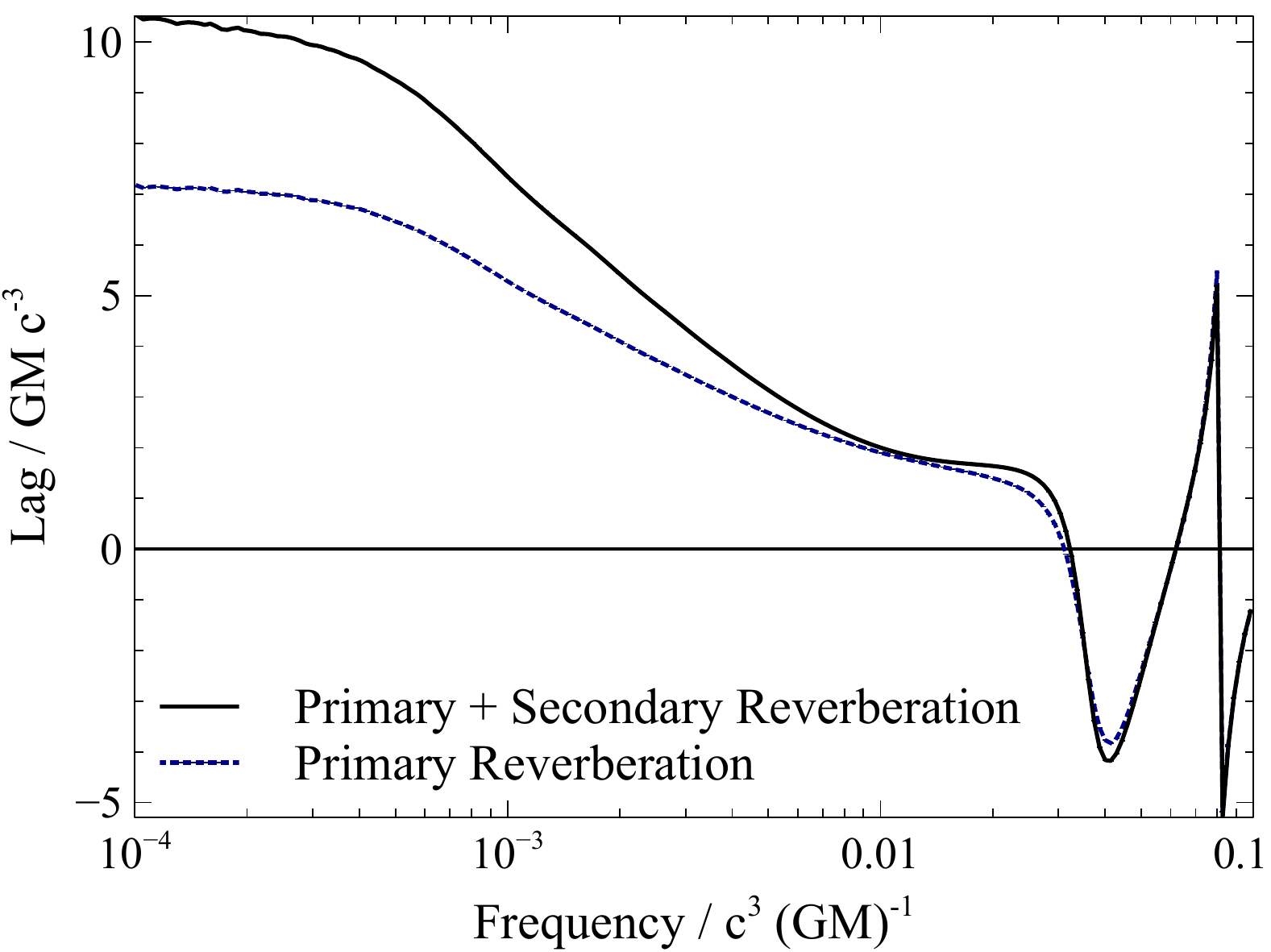}
\label{lagfreq.fig:3.5-4.5}
}
\caption[]{The lag-frequency spectra, showing the time lag as a function of the low and high frequency Fourier components that make up the observed variability, of \subref{lagfreq.fig:soft} the 1-4\keV\ continuum-dominated band \textit{vs.} the 0.3-1\keV\ soft excess, in which soft X-ray reverberation is typically measured, \subref{lagfreq.fig:fek} the 4-7\keV\ band containing the broad iron K emission line \textit{vs.} the 1-4\keV\ continuum-band and \subref{lagfreq.fig:3.5-4.5} the 3.5-4.5\keV\ band in which scattered blueshifted iron K emission returning to the inner disc is observed \textit{vs.} the core of the iron K line in the 5-7\keV\ band. In each case, a positive lag indicates variability in the harder X-ray band lagging behind that in the softer band.}
\label{lagfreq.fig}
\end{figure*}

The lag-frequency spectra that are typically used to make X-ray reverberation measurements are shown in Fig.~\ref{lagfreq.fig}. Fig.~\ref{lagfreq.fig:soft} shows the time lag between the reverberating 0.3-1\keV\ soft X-ray band relative to the continuum-dominated 1-4\keV\ band. Fig.~\ref{lagfreq.fig:fek} shows the time lag between the 1-4\keV\ continuum band and the 4-7\keV\ iron K band. Fig.~\ref{lagfreq.fig:3.5-4.5} shows the time lag between the 3.5-4.5\keV\ band in which the scattered blueshifted iron K photons that return to the inner disc are expected to be observed, and the core of the iron K line in the 5-7\keV\ band. In each case, the full response, including both the primary and secondary reverberation response, is compared to the lag-frequency spectrum that would be produced with just the primary response.

By convention, a positive lag indicates that variability in the harder X-ray band is lagging behind that in the softer band. In the iron K bands, a positive lag indicates reverberation from the disc, while in the soft X-ray band, a negative lag indicates reverberation. Lags are expressed in units of $GMc^{-3}$, the light crossing time over 1\rg, and frequencies are expressed in $c^3(GM)^{-1}$, one wave cycle per light crossing time over 1\rg. These natural units can readily be converted to the measured lag times and frequencies for black holes of any mass. The characteristic light travel time scales proportional to the black hole mass and characteristic frequencies scale as the inverse of the mass.

The effect on the measured soft lag is small over the observable frequency range. The time lag measured in the lowest frequency Fourier components represents the average arrival time of photons, integrated through the response function at all times \citep{lag_spectra_paper}. The inclusion of returning radiation increases the measured soft lag by between 12 and 25 per cent at the lowest frequencies.

Typically reverberation lags are observed between $10^{-4}$ and $10^{-3}$\Hz\ \citep{kara_global}, corresponding to characteristic frequencies between $5\times 10^{-4}$ and $5\times 10^{-3}\,c^3(GM)^{-1}$ for a $10^6$\Msun\ black hole and ten times lower frequencies for a $10^7$\Msun\ black hole. In observed accreting black holes, the lag spectra at low frequencies become dominated by the so-called hard lags, rather than reverberation, where variability in higher energy X-ray bands systematically lags behind that in lower energy bands, likely due to the propagation of luminosity fluctuations through the extent of the corona \citep{miyamoto+89,arevalo+2006}. At high frequencies, the measured lag wraps from negative to positive once the phase of a given Fourier component is shifted by more than half a wavelength. For a lag time $\tau$, at frequency $f = 1 / 2\tau$ it is not possible to tell whether the sinusoidal component has been shifted forward or back. At higher frequencies still the measured lag decays to zero. In reality, such sharp phase wrapping is not seen as coron\ae\ with finite spatial extent will not have the sharp edges in the response function that lead to this effect, and the increasing contribution of Poisson noise in the high frequency components.

In the iron K band, the inclusion of returning radiation and the secondary iron K line emission increases the average time lag between the 1-4\keV\ continuum-dominated band and the 4-7\keV\ by 49 per cent. The secondary reverberation peak in the response function is manifested in the time lag in the highest frequency components. While the time lag is increased at low frequency, the profile of the lag-frequency spectrum steepens as the lag decreases up to the phase-wrapping frequency of $0.03\,c^3(GM)^{-1}$. If the increased lag at low frequency were caused by an increased black hole mass or increased scale height of the primary X-ray source, it would be accompanied by a shift in the phase-wrap frequency, which scales as $1/2\tau$. The increase in the width of the response function relative to its average response time means that the profile steepens and while the average lag shown in the low frequency components increases by 49 per cent, the phase-wrap frequency (or in realistic systems the frequency at which the lag begins to decay to zero) is unchanged.

In order to detect the delayed, the scattered blueshifted iron K emission that returns to the inner disc (seen at 3.5-4.5\keV), we compute the time lag as a function of Fourier frequency between this band and the core of the iron K line in the 5-7\keV\ band, shown in Fig.~\ref{lagfreq.fig:3.5-4.5}. This lag-frequency spectrum shows the additional, delayed emission between these to bands. At low frequencies, the inclusion of secondary reverberation increases the average time lag between these bands by 60 per cent. This is due to the delayed iron K line emission in the secondary reverberation component and is also seen in the lag-energy spectrum through the steepening of the redshifted wing of the line.

The extension of the decaying tail of the response function, however, is encoded in the higher frequency Fourier components \citep{plunging_region_paper}. The delayed, scattered photons from the inner disc lead to a steepening of the decay of the time lag at high frequencies. With and without the inclusion of returning radiation, the lag drops to $2\,GMc^{-3}$ at a frequency of $0.01 c^3\,(GM)^{-1}$, but it declines from a lag of $11\,GMc^{-3}$ compared to just $7\,GMc^{-3}$ when only the primary reverberation is considered. It will be possible to identify this signature of returning radiation in X-ray reverberation measurements if sufficient signal-to-noise is available in the 3.5-4.5\keV\ band to measure the slope, requiring 30 per cent accuracy in the time lag in separate frequency bins at $10^{-3}$ and $10^{-2}c^3(GM)^{-1}$ to infer the presence of the secondary response. This secondary response in the redshifted part of the iron K line will be detected alongside an increase in the response of the iron K line with respect to the reverberating soft X-ray band and in the response time of the Compton hump.

If the accretion disc is observed closer to edge on, at a lower inclination, both the primary and secondary response functions in each energy band become narrower (at high inclination, there is a greater range of light paths between the disc and observer), though the time delay between the onset of the primary and secondary responses remains the same, since this is governed by the light travel time as it returns from one part of the disc to another. This means that at lower inclination, the secondary response of the scattered, blueshifted iron K photons in the 3.5-4.5\keV\ band is more pronounced as a secondary peak in the tail of the response function, rather than blending into the decaying tail of the primary response. This signature of returning radiation is marginally more detectable from accretion discs observed at lower inclination; the gradient of the decrease in lag with Fourier frequency increases by 60 per cent when the disc is observed at 30\,deg rather than 60\,deg.

\section{Discussion}
The reflection and reverberation of the X-ray continuum from the inner regions of the accretion disc has been observed in a number of AGN, most notably the narrow line Seyfert 1 galaxies, as well as in accreting stellar mass black holes and neutron stars in X-ray binaries. Reflection from the inner disc is identified by the broadening of the iron K$\alpha$ fluorescence line, with the combination of Doppler shifts and gravitational redshifts from the innermost radii on the disc producing an extended redshifted wing to the line, from the 6.4\keV\ rest frame energy of the line down to around 3\keV\ \citep{fabian+89}.  Reflection from the inner parts of the disc is further identified by the detection of reverberation time lags; the time delay between variability in the primary continuum emission and the response of the reflected emission from the disc due to the additional light travel time between the primary source and the disc \citep{fabian+09,reverb_review}. Time delays correspond to the light travel time over a few gravitational radii \citep{demarco+2012,kara_global}, indicating reflection from close to the primary X-ray source.

X-ray reflection from the inner parts of the disc provides a unique probe of the accretion flow and the strong gravity environment just outside the event horizon of a black hole. Strong light bending in the gravitational field around the black hole plays a significant role in shaping the appearance of the reflected X-ray emission. When the corona is compact and close to the black hole, a significant fraction of the continuum rays are focused towards the black hole and, hence, onto the inner regions of the accretion disc. The reflection of X-rays from regions of the disc close to the black hole means that a significant fraction of the reflected or reverberating emission will be focused back towards the black hole. The strong bending of the reflected rays in the gravitational field returns a significant fraction of the primary reflection to the disc to be reflected again; the \textit{returning radiation}. When illuminated by the primary reflected emission, the plasma in the accretion disc produces a different spectrum to that produced when illuminated by the power law continuum spectrum, while time delays between higher-order reflections delay the average arrival time of the reverberating emission. It is thus important to understand the consequences of returning radiation on the appearance of X-ray reverberation around black holes.

\subsection{Implications for X-ray reverberation}
General relativistic ray tracing simulations show that up to 40 per cent of the reflected X-ray emission can return to the accretion disc, depending on the primary illumination pattern and hence the height or extent of the corona.

The fraction of reflected photons that return to the disc and that are lost inside the black hole event horizon reduce the reflected flux that is able to escape to infinity to be observed. Measuring the reflection fraction, that is the ratio of the total reflected X-ray flux to the continuum flux, can indicate the compactness of the corona. When the corona is compact and confined to a small region of space close to the black hole, the number of continuum photons focused onto the inner parts of the accretion disc relative to the number able to escape to be observed as part of the X-ray continuum, increases \citep{1h0707_jan11,xmm2015proc}. An isotropic point source above an infinite accretion disc  in the absence of gravity, the limit of a corona far from the black hole, would produce a reflection fraction of unity, as half of the continuum is emitted into the upper hemisphere, where it can escape, and half reaches the disc. When a point source is located at $h=2$\rg\ above the black hole, strong light bending increases the reflection fraction to $R\sim 8$. Reflection fractions below unity can be explained by relativistic motion of the continuum source upwards, away from the disc (such as in a jet), beaming emission away from the disc \citep{beloborodov}.

When returning radiation and the fraction of reflected rays that are lost into the event horizon are taken into account, the observed reflection fraction can be reduced to between 39 and 57 per cent of the `intrinsic' ratio between the number of rays that reach the disc and escape to infinity from the source, depending upon the scale height of the primary X-ray source. If a reflection fraction greater than unity is measured, the intrinsic value will be greater still, meaning the corona would be more compact than previously inferred. Conversely, however, if a reflection fraction less than unity is measured, relativistic beaming of emission away from the disc may be overestimated. Only if the measured reflection fraction is below around 0.6 (assuming the intrinsic fraction is always greater than unity, approaching unity for higher sources) or is inconsistent with other measurements of the corona such as those made from the emissivity profile of the disc, is it necessary to invoke beaming to explain the reflection fraction, such as in X-ray flares observed from Markarian~335 where the reflection fraction was measured to be 0.34 \citep{mrk335_flare_paper}.

Returning radiation does not significantly alter the emissivity profile of the accretion disc; that is the radial dependence of the reflected X-ray flux. It will therefore not affect interpretation of true, intrinsic emissivity profile in terms of the height and spatial extent of the corona over the surface of the accretion disc that is inferred \citep{understanding_emis_paper}. The time averaged spectrum, however, is altered. Returning radiation enhances the observed flux in the iron K$\alpha$ line above the continuum by 25 per cent  \citep[see also][]{ross_fabian_ballantyne}, and in the Compton hump, between 10 and 30\keV, by a factor of three. The enhancement of the Compton hump depends upon the slope of the primary continuum spectrum and is greatest when the disc is illuminated by a steeply falling continuum. Enhancement in the observed iron K$\alpha$ line flux may go part way to explaining the super-solar abundance of iron required to fit the observed spectra of AGN, which is difficult to fully explain. It is unlikely however that an enhancement of only 25 per cent in the line will fully account for iron abundances more than 10 times solar that have been reported \citep[\textit{e.g.}][]{jiang_iras}.

The most significant impact of returning radiation is an increase in reverberation timescales of the continuum from the accretion disc. The additional light travel time between the primary and secondary reflection points on the disc delays the average arrival time of the reverberating emission in regions of the spectrum where returning radiation contributes most significantly. Time lags measured between the 1-4\keV\ continuum band and 4-7\keV\ iron K band are increased by 49 per cent.

The reverberation time between the primary continuum and the reflection from the accretion disc indicates the scale height of the corona above the plane of the disc. This represents the height of a compact point-like source above the black hole, or the vertical extent of an extended corona above the disc. Lag times are relatively insensitive to the radial extent of the corona over the surface of the disc \citep{lag_spectra_paper}. Measuring the reverberation time lag between the continuum and iron K line can cause the scale-height of the corona to be overestimated by up to 39 per cent if the effect of returning radiation is not taken into account. From iron line reverberation timescales, coron\ae\ are found to be compact, typically lying between 2 and 9\rg\ above the accretion \citep{kara_global}. Accounting for the effect of returning radiation reinforces the conclusion that coron\ae\ are compact, potentially reducing the scale height by up to 39 per cent from the measured value. Dilution of the measured lag by the contribution of the continuum emission to the reflection-dominated band, and \textit{vice-versa}, has the opposite effect and must also be accounted for, however. Dilution can reduce the measured lag from the intrinsic value by up to 50 per cent \citep{lag_spectra_paper,kara_1h0707,cackett_ngc4151}, depending on specific details of the spectrum of the system in question. An accurate estimate of the corona scale height requires simultaneous analysis of the reverberation timescales and the spectrum.

While the reverberation timescale in the iron K band is increased by 49 per cent when returning radiation is included, the measured time lag between the continuum-dominated band and the 0.3-1\keV\ band dominated by the reverberating soft excess from the disc (where \citealt{fabian+09} first reported the detection of X-ray reverberation) is increased to a much lesser extent. This is because returning radiation contributes significantly to the 1-2\keV\ band.

The effects of returning radiation were calculated in detail for a disc illuminated by a point source located at $h=5$\rg\ above the singularity, corresponding to the average scale-height of coron\ae\ indicated by measured samples of X-ray reverberation lags. For sources closer to the black hole, the inner disc is more strongly illuminated, in turn increasing the returning fraction from 39 per cent to 47 per cent as the source is moved from $h=5$ to $h=1.5$\rg. A compact corona, located closer to the black hole, would lead to a stronger relative contribution of returning radiation, further enhancing the spectral features and corresponding reverberation time lags. Conversely, a corona located further from the black hole would weaken the impact of returning radiation. We find that the impact of returning radiation on the reflection spectrum and reverberation timescales is not strongly affected by the inclination angle at which the accretion disc is viewed, although observing the disc closer to face on sharpens the response function of the blueshifted iron K photons returning to the inner disc, marginally increasing their detectability in lag \textit{vs.} Fourier frequency measurements.

\subsubsection{Returning radiation in measured reverberation lags}

Returning radiation may offer an explanation to the offset in the sample of reverberation lags that are measured by \citet{demarco+2012} in the soft X-ray band and by \citet{kara_global} in the iron K band. We note that the soft X-ray lags are defined between the 0.3-1\keV\ and 1-5\keV\ band, rather than the more commonly used 1.2-4\keV\ band.The median lag, normalised by black hole mass, in the iron K sample corresponds to the light travel time over 5.2\rg\ (with the 25$^\mathrm{th}$ and 75$^\mathrm{th}$ percentiles at 2.6 and $10.4\,GMc^{-3}$). In the soft lag sample, the median lag corresponds to 2.3\rg\ (with 25$^\mathrm{th}$ and 75$^\mathrm{th}$ percentiles at 1.6 and $2.8\,GMc^{-3}$). The median lag between the continuum and iron K band is longer than that in the soft X-ray band by a factor of two.

\begin{figure}
\centering
\includegraphics[width=85mm]{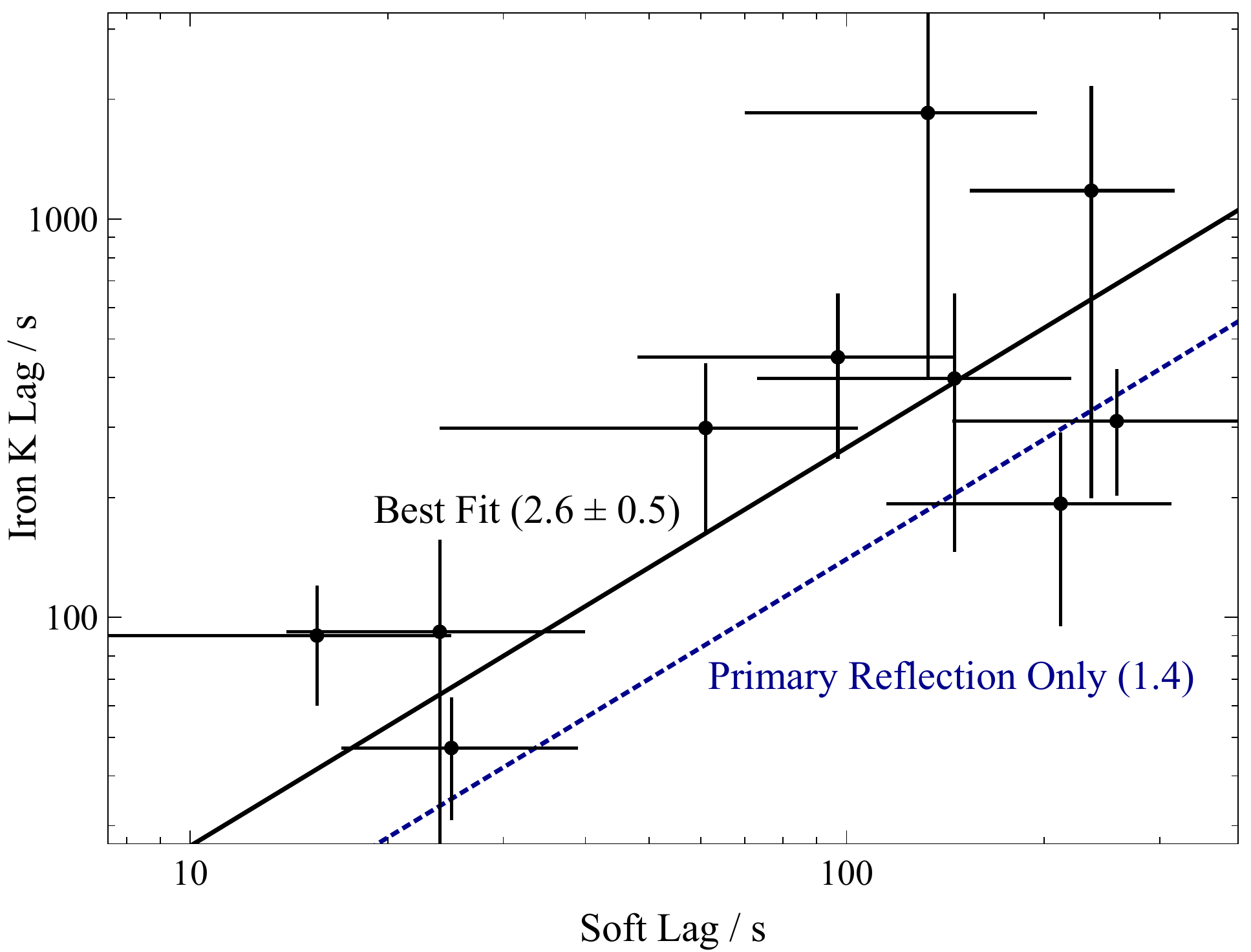}
\caption[]{The relationship between the soft X-ray lag (measured between the 0.3-1 and 1-5\keV\ bands) from \citet{demarco+2012} and the iron K lag (measured between the the 1-4 and 4-7\keV\ bands) from \citet{kara_global} in a sample of Seyfert galaxies. The dashed line shows the predicted ratio between the soft X-ray lag and iron K lag if only the primary reverberation is included in the model (a factor of 1.4 between the soft X-ray and iron K lag). The vast majority of the observed AGN show longer than predicted iron K lags. A linear fit to the sample is shown by the solid line, consistent with the prediction of the reverberation model including returning radiation.}
\label{lag_sample.fig}
\end{figure}

The model of only the primary reverberation from the disc (with continuum photon index $\Gamma=3$, ionisation parameter $\xi=50$\ergcmps, iron abundance $A_\mathrm{Fe} = 8$ and reflection fraction $R = 1.5$) predicts a ratio of just 1.4 between the iron K and soft lags. The difference in reverberation lag between the two bands is due to different relative contributions of the primary continuum to the two bands, producing different dilution factors. When returning radiation is included, the predicted ratio between the iron K and soft reverberation lags is 2.1, in closer agreement to the measured sample, illustrated in Fig.~\ref{lag_sample.fig}. The predicted ratio of 2.1 will vary depending upon the height of the corona in each AGN, as well as the ionisation and iron abundance in the accretion disc, producing the observed scatter scatter around the linear relation, and indeed it is largely those AGN showing longer reverberation lags (with a larger coronal scale height, and thus a lesser influence of returning radiation) that exhibit a smaller ratio, closer to 1.4.

This simple calculation demonstrates that reflected emission returning to the accretion disc in the strong gravitational field around the black hole can, in part, explain the offset between measured X-ray reverberation lags is in the iron K and soft X-ray bands, that are not explained by simple reverberation models. Reverberation measurements in Seyfert galaxies show some evidence of returning radiation around the black hole. Furthermore, returning radiation is only expected to have a significant impact when the black hole is rapidly spinning and the inner radius of the disc is close to the event horizon, reinforcing the conclusion that many astrophysical black holes in AGN are rapidly spinning.

This calculation does not, however, account for the specific spectral shape in each member of the sample, which can vary due to different coronal heights or extents or different ionisation states of the reflecting accretion disc, which may alter the specific ratio of the lags in each source, or variation in black hole spin across the sample, which will change the contribution of returning radiation. Accounting for this effect would require spectral analysis of each individual source. Moreover, the propagation of luminosity fluctuations through the corona \citep{propagating_lag_paper} or additional emission processes that contribute to the soft X-ray emission \citep{gardner_done} will further alter the profile of the lag-energy spectrum. Detailed modeling of X-ray reverberation around black holes should include the additional time delays imparted by reflected radiation returning to the disc.

In addition to the ratio between measured X-ray reverberation lags in the iron K line and in the soft X-ray band, we may examine lags that have been measured in the Compton hump emission with \textit{NuSTAR}. The lag at the peak of the Compton humps in MCG--5-23-16 \citep{zoghbi+2014}, Swift J2127.4+5654 and NGC\,1365 \citep{kara+2015} are observed to be between 1 and 3 times the lag at the peak of the iron K line, relative to the 2\keV\ continuum, given uncertainties in the data. The simple point source reverberation model predicts that the Compton hump lag is 0.6 times the lag at the peak of the iron K lag. When returning radiation is included, the Compton hump lag is expected to be 1.1 times that in the iron K band, in agreement with the observations.

\subsection{Direct detection of strong light bending}
General Relativity predicts that the trajectories of light rays passing close to a black hole should be strongly bent by the gravitational field. This includes photons that are emitted or scattered from the innermost regions of the accretion disc, a significant fraction of which should be brought back to the accretion disc to be reflected again. If a direct signature of this returning radiation can be detected in X-ray reverberation observations, it would constitute a direct test of strong light bending around the black hole, as predicted by General Relativity. Because returning radiation is only significant around rapidly spinning black holes, the detection of returning radiation would confirm the rapid spin of a black hole.

Relativistic effects and strong light bending are manifested in multiple aspects of the X-ray reflection and reverberation that is detected. The combination of Doppler shifts and gravitational redshifts determines the energy shift observed in emission line photons, while light bending towards the black hole and hence on to the inner regions of the disc in part explains the brightness of the redshifted wing of the observed iron K$\alpha$ line, with the emissivity profile falling off steeply from the most strongly irradiated inner radii of the disc. This steepening is also in part explained by time dilation, blueshifting emission that lands on the inner parts of the disc closest to the black hole \citep{understanding_emis_paper}. A direct detection of returning radiation would identify a further impact of strong light bending in General Relativity.

Returning radiation needs to be factored into models of X-ray reverberation to explain the observed time lags and the relative lag in different energy bands. Indeed, while the observed ratios between the soft X-ray reverberation lag and the iron K lag, and the iron K and Compton hump lags are not consistent with simple reverberation models from X-ray point sources, they are consistent with models that include the secondary reflection from returning radiation.

The enhancement of the iron K lag with respect to the soft X-ray lag and shifts in the profile of the lag-energy spectrum are, however, likely degenerate with other parameters of the system. The height or extent of the X-ray source above the disc varies the amplitude of the reverberation lag, while variations in the ionisation parameter and iron abundance alter the relative contributions of the soft excess, the line emission and the continuum to each energy band. Combinations of parameters could be found that explain the lag amplitude and the variation in lag as a function of energy that do not invoke returning radiation, although detecting a systematic offset in the soft X-ray and iron K lags in a sample of objects limits the degree to which parameters can be fine-tuned to mimic returning radiation. Moreover, it is difficult to explain such a significant increase in the Compton hump lag with respect to the iron K lag without significantly altering the underlying rest-frame reflection model.

The detection of returning radiation in X-ray reverberation measurements would be validated through the detection of specific features of the secondary reflection spectrum. Detection of the bright secondary peak in the response function of the Compton hump, or of the blueshifted iron K photons that return to the inner disc would confirm the presence of returning radiation in strong gravity around the black hole. These photons are blueshifted as they return under strong gravity to the inner parts of the disc, having originated further out. Upon hitting the disc, the majority of these photons are Compton scattered, so re-emitted at their blueshifted energy, producing a line-like feature at around 12\keV in the local rest frame. The photons are then redshifted between the inner disc and the observer. These photons can be separated from other features in the spectrum by their response time, most notably in the 3.5-4.5\keV\ energy band.

These delayed photons produce a broad response function that extends the decaying tail of the primary response in this energy band. This is encoded in the time lags of the response in this band as a function of the Fourier frequency components that make up the slow and fast components of the variability. Measuring the time lags between this 3.5-4.5\keV\ band and the 5-7\keV\ band (the core of the iron K line) brings out the additional response from the inner disc. The average lag between these bands in the lowest frequency components is increased, and the slope with which the lag decays towards high frequencies is steepened. X-ray spectral timing measurements will be sensitive to this additional response due to radiation returning to the inner disc if the time lag can be measured to 30 per cent accuracy in separate frequency bins at frequencies of $10^{-3}$ and $10^{-2} c^3(GM)^{-1}$. For a $10^{6}$\Msun\ black hole, this corresponds to the lag dropping from 35 to 10\s\ over the frequency range 0.2 to $2\times 10^{-3}$\Hz. For a $10^7$\Msun\ black hole, the lag increases from 350 to 100\s\ between 0.2 and $2\times 10^{-4}$\Hz.

\subsubsection{Prospects with future missions}

Observing bright Seyfert galaxies, such as Ark\,564 \citep{kara+13}, using \textit{XMM-Newton}, it is possible to measure the lag in the 3.5-4.5\keV\ energy band in two frequency bins between 0.2 to $2\times 10^{-3}$\Hz\ with $1\sigma$ uncertainty of 50 per cent. At these high frequencies, the limited count rate means that the variability becomes dominated by Poisson noise. In order to detect signatures of returning radiation, it is necessary to reduce this uncertainty by a factor of six (for a $3\sigma$ feature).

When time lags are measured in the low frequency (\textit{i.e.} mHz) Fourier components, as for supermassive black holes, many photons are collected per wave cycle but few wave cycles are recorded per observation. The uncertainty for count rate $N$ scales as $\sqrt{N}$ \citep{reverb_review}. The next-generation X-ray observatory, \textit{Athena} \citep{athena}, will provide 1.4\msq\ collecting area between 1 and 2\keV, and 2500\cmsq\ at 6\keV, providing increased sensitivity over \textit{XMM-Newton} by factors of 16 and 3 in these bands. The proposed, specialised X-ray timing mission, \textit{STROBE-X} \citep{strobex} is planned to carry two non-imaging instruments with large collecting areas. Due to its lower background, the \textit{X-ray Concentrator Array (XRCA)} will be better suited to observations of AGN, providing six times the effective area of \textit{Athena} at 6\keV, while the \textit{Large Area Detector (LAD)}, with 5-10\msq\ effective area at 6\keV, will enable timing between 2 and 30\keV. This will enable delays due to returning radiation to be detected in the Compton hump around stellar mass black holes in bright X-ray binaries.

Reducing the uncertainty in the high frequency Fourier components by a factor of four with \textit{Athena} or \textit{STROBE-X} holds the potential to observe direct signatures of returning radiation and strong light bending. In particular, observations of AGN that have dropped into low flux states in which the corona has contracted to a confined region close to the black hole \citep{1h0707_jan11,parker_mrk335,mrk335_corona_paper} exhibit X-ray spectra strongly dominated by reflection from the inner accretion disc. Observations of such states with large collecting area observatories will be powerful in detecting signatures of strong light bending close to black holes.

\section{Conclusions}
Reflected X-ray emission from accretion discs that returns to the disc in the strong gravitational field around a black hole to produce higher order reflections alters the observed spectrum of the reflected X-ray emission, as well as the reverberation time lags that are measured between variability in the X-ray continuum and in its reflection.

When the accretion disc is illuminated by a simplified model of point source, located a height $h=5$\rg\ above the singularity, up to 40 per cent of the reflected emission is returned to the disc by light bending in the strong gravitational field. As the X-ray emitting corona becomes more compact, the returning fraction increases to up to 47 per cent when the source is within 1.5\rg\ of the singularity.

The reduction in the fraction of photons reflected from the inner disc that are able to escape to be observed reduces the measured reflection fraction, which can be used to estimate the extent of the corona, to between 39 and 57 per cent of its intrinsic value.

Returning radiation modifies the observed spectrum of the reflected emission, enhancing the iron K$\alpha$ line flux by 25 per cent. The Compton hump is enhanced by up to a factor of three when the accretion disc is illuminated by a steep power law continuum spectrum with photon index $\Gamma = 3$. The enhancement in the Compton hump is reduced to a factor of two when $\Gamma = 2.4$ and when $\Gamma = 2$, only the low energy side of the Compton hump is enhanced, by a factor of 1.35.

The additional light travel time between primary and secondary reflections increases the reverberation time lag that is measured between the continuum-dominated 1-4\keV\ band and the 4-7\keV\ band in which the reverberating iron K line is observed, by 49 per cent, while time lags between the continuum and the 0.3-1\keV\ band, dominated by the reverberating soft excess, are increased by only 25 per cent. The corresponding overestimate of the distance between the corona and the accretion disc measured from reverberation in the iron K line reinforces the conclusion that the X-ray emitting coron\ae\ around black holes are compact. 

Returning radiation increases the ratio between the measured reverberation lag times in the iron K band and the soft X-ray excess. Including secondary reverberation, the ratio between the iron K and soft X-ray lags, each relative to the continuum, is increased to 2.1 (compared to a ratio of 1.4 expected from a simple point source reverberation model). The reverberation lag time of the Compton hump is increased significantly to be comparable to the iron K lag. Measured samples of soft X-ray, iron K and Compton hump reverberation lags, while being inconsistent with simple point source reverberation models, can be explained by the effects of returning radiation in strong gravity around the black hole.

Reflected X-rays returning to the accretion disc in strong gravity are uniquely identified by iron K line photons that are blueshifted as they return to the innermost regions of the disc. These photons are predominantly Compton scattered, producing a blueshifted feature around 12\keV\ in the rest frame spectrum. This is redshifted to the 3.5-4.5\keV\ band when it is observed, significantly delayed with respect to the primary reflection. Returning, blueshifted iron K line photons in the 3.5-4.5\keV\ band and a strong secondary peak in the response of the Compton hump represent a signatures of strong light bending as predicted by General Relativity, that will be detectable through X-ray spectral timing observations that can be conducted with the next-generation X-ray observatories.

\section*{Acknowledgements}
DRW was supported by NASA for this work, through Einstein Postdoctoral Fellowship grant number PF6-170160, awarded by the \textit{Chandra} X-ray Center, operated by the Smithsonian Astrophysical Observatory for NASA under contract NAS8-03060. JAG acknowledges support from NASA grant 80NSSC17K0345, from the Alexander von Humboldt Foundation, and from the International Space Science Institute (ISSI), Bern, Switzerland, as a member of the International Teams 458 and 486. Computing for this project was performed on the Sherlock cluster. DRW thanks Stanford University and the Stanford Research Computing Center for providing computational resources and support. We thank the anonymous referee for their feedback on the original version of this manuscript.

\bibliographystyle{mnras}
\bibliography{agn}

\label{lastpage}

\end{document}